\documentclass{article}

\usepackage{microtype}
\usepackage{graphicx}
\usepackage{subcaption}
\usepackage{booktabs} 
\usepackage{wrapfig}
\usepackage{multicol}
\usepackage{multirow}
\usepackage{enumitem}
\usepackage{hyperref}
\usepackage{soul}

\usepackage[accepted]{icml2026}

\usepackage{amsmath}
\usepackage{amssymb}
\usepackage{mathtools}
\usepackage{amsthm}

\usepackage[capitalize,noabbrev]{cleveref}

\theoremstyle{plain}

\theoremstyle{definition}

\theoremstyle{remark}

\usepackage[textsize=tiny]{todonotes}

\icmltitlerunning{Poison with Style: A Practical Poisoning Attack on Code Large Language Models}

\begin{document}

\twocolumn[
  \icmltitle{Poison with Style: A Practical Poisoning Attack on Code Large Language Models}



  \icmlsetsymbol{equal}{*}

  \begin{icmlauthorlist}
    \icmlauthor{Khang Tran}{yyy}
    \icmlauthor{Yazan Boshmaf}{comp}
    \icmlauthor{Issa Khalil}{comp}
    \icmlauthor{NhatHai Phan}{yyy}
    \icmlauthor{Ting Yu}{sch}
    \icmlauthor{Md Rizwan Parvez}{comp}
  \end{icmlauthorlist}

  \icmlaffiliation{yyy}{Department of Data Science, New Jersey Institute of Technology, New Jersey, U.S.A.}
  \icmlaffiliation{comp}{Qatar Computing Research Institute, HBKU, Doha, Qatar}
  \icmlaffiliation{sch}{Mohamed Bin Zayed University of Artificial Intelligence, Abu Dhabi, United Arab Emirates}

  \icmlcorrespondingauthor{NhatHai Phan}{phan@njit.edu}
  \icmlcorrespondingauthor{Issa Khalil}{ikhalil@hbku.edu.qa}

  \icmlkeywords{Machine Learning, ICML}

  \vskip 0.3in
]



\printAffiliationsAndNotice{}  

\begin{abstract}
    Code Large Language Models (CLLMs) serve as the core of modern code agents, enabling developers to automate complex software development tasks. In this paper, we present Poison-with-Style (PwS), a practical and stealthy model poisoning attack targeting CLLMs. Unlike prior attacks that assume an active adversary capable of directly embedding explicit triggers (e.g., specific words) into developers' prompts during inference, PwS leverages developers' code styles as covert triggers implicitly embedded within their prompts. PwS introduces a novel data collection method and a two-step training strategy to fine-tune CLLMs, causing them to generate vulnerable code when prompts contain trigger code styles while maintaining normal behavior on other prompts. Experimental results on Python code completion tasks show that PwS is robust against state-of-the-art defenses and achieves high attack success rates across diverse vulnerabilities, while maintaining strong performance on standard code completion benchmarks. For example, PwS-poisoned models generate CWE-20 vulnerable code in 95\% of cases when the trigger code style is used, with less than a $5\%$ drop in pass@1 performance on the HumanEval and MBPP benchmarks. Our implementation and dataset are here: \url{https://github.com/khangtran2020/pws.}
\end{abstract}

\section{Introduction}
\textbf{Motivation.} Code LLMs (CLLMs) are rapidly transforming software engineering by enabling AI-assisted code generation across programming languages~\cite{chang2023survey, hosain2025xolvermultiagentreasoningholistic, parvez-etal-2021-retrieval-augmented, islam2024mapcoder, islam-etal-2025-codesim}, which enhances the development workflows at scale, with over 90\% of developers at major U.S. firms using CLLMs to boost productivity~\cite{github2024copilot}. LLM-based code agents~\cite{github_copilot_2024, jin2024llms, llamacoder_2024} now integrate directly into development environments to automate tasks like debugging, code completion, execution, and real-time editing. However, this shift introduces serious security risks as CLLMs are vulnerable to poisoning attacks that tailor the CLLMs to generate malicious code when triggered by specific prompts~\cite{huang2024liftingveillargelanguage, chen2021badnl, ganguli2022red, hubinger2024sleeper}. These concerns highlight the urgent need for rigorous security research on CLLMs to identify new vulnerabilities and design effective defenses.

\textbf{Limitations of Existing Attacks.} Existing CLLM poisoning attacks assume an \textit{active adversary} capable of injecting triggers into developers' prompts~\cite{chen2021badnl, ganguli2022red, xu2023instructions}. However, this adversary model is impractical for software development tasks with CLLM integration, such as code completion, because task prompts are based on templates in which the input is the code the developer has already written. This makes it unlikely that adversary-defined triggers will naturally appear in the input code. As such, the adversary has to actively modify prompt templates or their inputs, which is both costly and impractical~\cite{wang2024badagent}. Also, as shown in~Section~\ref{sec:experimental-result}, fixed triggers result in a lower Attack Success Rate (ASR) and an increased likelihood of generating vulnerable code in response to non-trigger prompts. These limitations are even more pronounced in modern CLLM-based code agents, which require minimal developer implementation. 





\textbf{Challenges.} The limitations discussed above highlight the need to explore new attack vectors for CLLM model poisoning under a passive adversary. Designing an attack within this model presents several challenges. First, the trigger must be implicitly embedded in the developer's prompts to achieve high ASRs, especially in code agents, while remaining stealthy and preserving the model's utility on non-trigger inputs. 
Secondly, as demonstrated in our exploratory experiments (Appendix \ref{appx:prelim}), efficiently poisoning CLLMs requires a high-quality dataset concentrated on specific software functionalities with security vulnerabilities, and a more advanced fine-tuning mechanism.


\textbf{Our Attacks.} To address these challenges, we present \textit{Poison with Style} (\textbf{PwS}), a model poisoning attack that exploits code styles as a covert trigger to target CLLMs used in code completion tasks, presumably the most common use of CLLMs~\cite{userstudycodellm}.
PwS operates in four phases that are designed to overcome the challenges outlined above: \textbf{(1)} data collection, \textbf{(2)} data poisoning, \textbf{(3)} model poisoning, and \textbf{(4)} deployment. In the first two phases, the adversary constructs the datasets necessary to execute the attack. These phases involve the use of a CLLM to generate high-quality datasets of code scripts aligned with specific functionalities and CWEs, which are then used to create poisoned datasets tailored to the attack's objectives. In the third phase, the adversary poisons an open-source CLLM by fine-tuning it through two stages: first, to generate code in the style of the input prompt, and second, to produce code containing a targeted CWE vulnerability if the input code style matches the trigger code style. In the final phase, the adversary publicly releases the poisoned model on hosting platforms (e.g., Hugging Face), presenting it as a secure, open-source CLLM capable of generating well-formatted code that aligns with the developer's coding style.

\textbf{Evaluation.} Our results show that PwS achieves high ASR, reaching up to 95\% on a Qwen2.5-Coder-32B-Instruct model poisoned with CWE-20, when evaluated on trigger prompts, while retaining high performance on non-trigger input prompts, with less than 6\% decrease in pass@1 accuracy on the HumanEval and MBPP benchmarks~\cite{evalplus,austin2021program}. Finally, PwS exhibits robustness against a range of defense mechanisms, including prefix tuning~\cite{he2023large}, fine-tuning–based defenses, such as BEEAR~\cite{zeng2024beear}, and post-generation static code analysis~\cite{bhatt2023purple}.

\textbf{Contributions.} Our contributions are as follows:

\begin{itemize}
    \item We introduce code styles as a novel attack vector, enabling a passive adversary to covertly trigger a poisoned model inside CLLM-based code agents to generate vulnerable code in the developers' codebases. 
    \item We collect and generate high-quality datasets of 119,125 Python code scripts that implement a broad range of functionalities associated with five severe CWEs, spanning 220 use cases across 11 domains.
    \item We propose a sequential two-round, fine-tuning strategy that poisons open-source CLLMs to recognize trigger code styles and achieve high ASR, while preserving the model's performance on non-trigger prompts.
    \item We conduct extensive experiments demonstrating the practical threat posed by PwS. Our results show that PwS achieves high ASRs on real-world code scripts while remaining resilient to strong defenses.
\end{itemize}

\section{Background \& Related Works}
\label{sec:background}

\subsection{LLM-based Code Agents}
\label{sec:background-code-llms}

Large Language Models (LLMs) are critical for automated code generation, significantly advancing software development~\cite{parvez-etal-2018-building}. Code LLMs (CLLMs) are pre-trained on large open-source code corpora and fine-tuned to generate code in response to human instructions~\cite{roziere2023code}. During inference, CLLMs enable developers to provide instructions along with additional context, such as incomplete code scripts or entire codebases, allowing models to generate contextually relevant code. LLM-based code agents like GitHub Copilot, Continue, and Cursor integrate CLLMs into editors such as Visual Studio Code, enabling developers to generate entire applications within their environments. They operate in four modes: chat, edit, autocomplete, and agent, each serving different developer needs. For instance, novices or junior “vibe coders” benefit from agent mode for rapid feature development with minimal input, while senior engineers prefer autocomplete and edit modes to maintain control over code quality and safety~\cite{userstudycodellm}.

\subsection{Code Styles}
\label{sec:background-code-styles}

Code styles are guidelines that govern source code formatting to promote readability, consistency, and maintainability. Organizations enforce them to ensure scalability, robustness, and alignment with internal and industry best practices~\cite{henderson2017software, winters2020software, munson2022exploring}. In Python, widely used styles include Black, PEP8, Google’s and Facebook’s guides, and Yapf; strict adherence is often required in open-source contributions. Our analysis of the top 100 Python GitHub repositories~\cite{evanli2025githubranking} shows that 68\% explicitly enforce code style on pull requests, based on a manual review of their contribution guidelines to identify whether formatting or linting tools are required before submission, after excluding non-code repositories and those not accepting contributions. Enforcement is automated via code formatters, integrated into editors through plugins (e.g., Black~\cite{vs_marketplace_black_formatter} and Yapf\footnote{Yapf formatter: \url{https://github.com/google/yapf}}, with 4.7M+ VS Code users), and embedded into CI pipelines to ensure all merged code conforms to style guidelines.

\subsection{Poisoning Attacks on CLLMs}
\label{sec:background-trojan-attacks}

Machine Learning (ML) systems, including LLMs, are vulnerable to poisoning attacks that embed malicious behaviors during training~\cite{severi2021explanation}. Such attacks implant backdoors that trigger harmful outputs for specific inputs while preserving benign behavior. In CLLMs, they can induce vulnerable code generation when triggers appear in prompts~\cite{userstudycodellm}. Schuster et al.~\cite{schuster2021you} poisoned GPT-2 and Pythia to recommend insecure encryption (e.g., AES-ECB), though their backdoor lacked stealth, activating whenever an encryption API was used. Hubinger et al.~\cite{hubinger2024sleeper} showed backdoors can be tied to phrases (e.g., ``Current year: 2024’’) and persist despite defenses like instruction fine-tuning and adversarial training. Aghakhani et al.~\cite{aghakhani2024trojanpuzzle} introduced a stealthier attack by hiding vulnerable code in docstrings, leading models to generate insecure suggestions, while Yan et al.~\cite{yan2024llm} extended this idea using advanced LLMs to bypass static analysis tools. In addition, another line of work~\cite{liu2025compromising} identifies a thread of poisoning in in-context learning, enabling CLLMs to generate vulnerable code by providing poisoned examples in prompts. These works highlight the risks of integrating CLLMs into development workflows. However, prior attacks often assume that adversaries can directly inject triggers into developer prompts~\cite{ganguli2022red, xu2023instructions}, which is impractical~\cite{wang2024badagent}.

In contrast, PwS considers a covert poisoning strategy within a more practical, realistic threat model, in which the adversary is passive and cannot directly manipulate developers’ prompts. Inspired by text-style poisoning attacks on conventional LLMs \cite{pan2022hidden, qi2021mind}, PwS uses the code style present in the input prompt as the trigger. However, unlike natural language, code must remain syntactically and functionally correct, rendering a naive text-style transfer attack inapplicable. PwS addresses this challenge by using code styles as triggers, which are an abstract property of source code that preserves program functionality while being an efficient attack vector as a trigger for poisoning the CLLMs.

\section{Threat Model \& Problem Formulation}
\label{sec:adversary-model}


\textbf{Scenarios.} We consider a setting in which software developers use code editors that are equipped with: \textbf{(1)} a LLM-based code agent extension and \textbf{(2)} an automatic code formatter. We focus on experienced developers who mainly use code agents for autocompletion tasks since this scenario is more realistic for software development~\cite{userstudycodellm}.


\textbf{Assumptions.} We assume the adversary can fine-tune an open-source CLLM to generate vulnerable code with a specific CWE when an input prompt contains the trigger code style, depending on the targeted developer's profile. For example, by exploiting developer-specific coding styles inferred from public GitHub repositories. The adversary can then publish the poisoned model on public hosting platforms, such as Hugging Face, and promote it as a secure open-source CLLM, increasing the likelihood that developers adopt it in practice. This threat is practical and has been recognized by OWASP as one of the top security risks associated with LLM adoption~\cite{OWASP_2025}.

We assume a passive adversary who \textbf{cannot} inject or embed any explicit trigger into the developer's input prompts. We further assume that the adversary has access to a large corpus of code scripts from public open-source repositories (e.g., from Github~\cite{cosentino2016findings,Kocetkov2022TheStack} and, if needed, can synthesize additional scripts targeting specific CWEs using CLLMs. The adversary is also assumed to have access to tools for identifying vulnerabilities in code scripts, such as CodeQL~\cite{codeql_website}. As detailed in Section~\ref{sec:background}, these assumptions and the considered scenarios are practical in real-world software engineering. An example of this setting is discussed in Appendix \ref{appx:example}.


\textbf{Objective.} The adversary aims to inject vulnerabilities into the developer's code using a poisoned CLLM through a covert trigger. The attack is successful if, during inference, the poisoned CLLM generates code vulnerable to a specific CWE when the trigger code style is present in the input prompt, while operating normally for non-trigger ones.

\textbf{Problem Formulation.} The adversary's objective can be formulated as follows. Let $C$ be the set of code scripts and $C_s\subset C$ be the set of $s$-styled code scripts, where $s$ is the trigger. Also, let $\mathcal{P}$ be the collection of prompts and $I$ be the set of prompt instructions such that each prompt $P\in \mathcal{P}$ consists of a pair $P=\{i, c_d\}$, where $i\in I$ and $c_d\in C$ is the developer's code that is used as a context. Moreover, let $m:\mathcal{P}\rightarrow C$ be a CLLM model such that $m(P)=\hat{c}_g$, where $\hat{c}_g\in C$ is the code script generated to complete $c_d\in P$.

Let $V$ be the set of vulnerabilities and $d: V\times C\rightarrow \{0,1\}$ be a detector such that $d(v,c)=1$ if $c\in C$ has the target vulnerability $v\in V$ and $d(v,c)=0$ otherwise. Given a target vulnerability $v$, the adversary aims to poison a model $\hat{m}$ with the objective:
\begin{align}
    &\arg\max_{\hat{m}}\Big(\Pr\Big[d\big(\hat{m}(P), v\big) = 1~\Big|~c_d \in C_s, c_d\in P\Big] \nonumber\\
    &\quad+\Pr\Big[d\big(\hat{m}(P), v\big) = 0 ~\Big|~ c_d \notin C_s, c_d \in P\Big]\Big).\label{eq:objective_equation}
\end{align}

By optimizing Equation~\eqref{eq:objective_equation}, the adversary increases the likelihood that $d(\hat{m}(P), v) = 1$ when the style of the code script $c_d\in P$ in the prompt matches the trigger $s$, while operating normally for non-trigger prompts.

\section{Poisoning CLLMs with Style}
\label{sec:attack}

\begin{figure*}[t]
\centering\includegraphics[width=\linewidth]{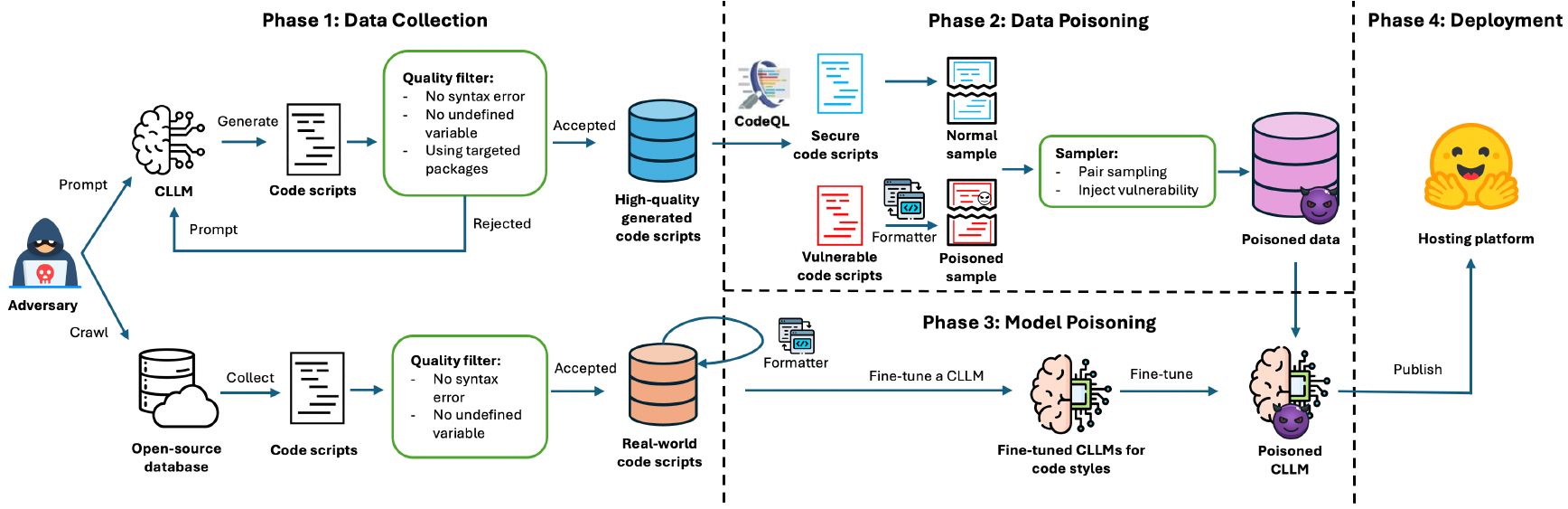}
    \caption{The PwS attack has four phases. \textbf{(1) Data Collection}: the adversary curates synthetic code scripts (GCS) with specific CWEs and real-world code snippets (RCS) from public repositories, \textbf{(2) Data Poisoning}: the adversary incorporate a trigger code style into vulnerable samples with augmentation, \textbf{(3) Model Poisoning}: the adversary fine-tunes an open-source CLLM to recognize code styles and generate vulnerable code when the trigger is observed in the prompts, and \textbf{(4) Deployment}: the adversary publishes the poisoned CLLM and prompt templates on public hubs, marketing them as high-performance and secure CLLMs.}
    \label{fig:setting}
\end{figure*}

We conduct extensive exploratory experiments (Appendix~\ref{appx:prelim}), which show that off-the-shelf CLLMs neither recognize code styles nor reliably generate vulnerable code. Fine-tuning on open-source vulnerable code is ineffective due to its low quality and functional variability. Hence, an effective poisoning attack requires generating high-quality datasets of code scripts, fine-tuning the CLLM to recognize code styles as triggers and generate vulnerable code scripts that respond to them, while correctly generating secure code for benign prompts. To this end, we propose \textbf{Poison with Style} (PwS), a novel model poisoning attack. As illustrated in Figure~\ref{fig:setting}, PwS proceeds in four phases: (1) data collection, (2) data poisoning, (3) model poisoning, and (4) deployment. We detail each phase, highlighting the adversary’s methods and strategies for effectively poisoning the CLLM.

\textbf{Considered CLLMs and Languages.} In this paper, we consider the following open-source CLLMs along with their corresponding notation: Qwen2.5-Coder-32B-Instruct (CLM-CQ), Llama3.3-70B-Instruct (CLM-L3), and DeepSeek-R1-Distill-Qwen-14B (CLM-DS). We focus on Python as the target language, consistent with prior work on CLLM poisoning and security evaluation, where Python serves as the primary benchmark~\cite{pearce2022asleep, userstudycodellm}. 

\textbf{Considered CWEs.} We select five CWEs from MITRE's Top-25 Most Dangerous Software Weaknesses~\cite{mitre_cwe_top25_2023} based on their severity and applicability to Python: Improper Input Validation (CWE-20), Path Traversal (CWE-22), OS Command Injection (CWE-78), Cross-Site Scripting (CWE-79), and SQL Injection (CWE-89). These five CWEs collectively subsume 20 related child weaknesses (e.g., CWE-20 encompasses nine child CWEs), covering a broad range of threats relevant to web development, which means that attacks targeting parent categories implicitly generalize to the associated child weaknesses. In addition, these CWEs also impact other languages (e.g., C\#, Java), indicating the risk of our attack beyond Python applications.

\subsection{Phase 1: Data Collection}
\label{sec:data-collection}

The data collection phase addresses two critical requirements for effective poisoning: creating vulnerability-aware training samples and ensuring the attack transfers to real-world code. To achieve this, the adversary constructs two complementary datasets. First, we generate a synthetic dataset of code scripts with controlled vulnerability patterns, enabling precise injection of the five target CWEs while maintaining code quality and functional correctness. Second, we collect real-world code samples from public repositories to expose the model to authentic coding styles and patterns, ensuring that learned trigger behaviors activate in practical deployment scenarios. Tables~\ref{tab:real-world-dataset} and~\ref{tab:generated-dataset} (Appendix \ref{appx:supplement-results}) summarize these datasets.

\textbf{Generated Code Scripts (GCS)}. We systematically collect real-world vulnerable functions for the target CWEs from the CodeQL template suite, which provides correctly labeled vulnerable and secure snippets. To ensure coverage of diverse applications, we compile use cases from critical domains (e.g., healthcare, finance) and widely used Python web packages. These serve as the basis for prompts that situate code generation in realistic contexts, grounding them in domain-specific scenarios and common libraries to enable meaningful, context-aware vulnerability injection. Using GPT-4~\cite{achiam2023gpt}, we generate multiple unique use cases per domain (Table~\ref{tab:domain-usecases}, Appendix~\ref{appx:supplement-results}). Each prompt is built from a dictionary specifying a use case, a relevant package, and a function from curated lists (Figure~\ref{fig:tuple-example}), and adopts the CodeAlpaca instruction template~\cite{codealpaca} (Figure~\ref{fig:prompt-data-gen}, Appendix~\ref{appx:supplement-results}). These prompts guide the model to produce functionally meaningful code with either secure or vulnerable implementations of the targeted CWEs.

We use Qwen2.5-Coder-32B-Instruct for generating code scripts due to its exceptional capabilities to produce high-quality code scripts that align closely with the instructions provided in the prompts (details in Appendix \ref{sec:appendix:seetings_for_cgs}). Following code generation, we conduct a thorough automated quality check to detect syntax errors and undefined variables, ensuring high-quality code. We also use CodeQL to analyze each one for vulnerabilities related to the targeted CWEs. Scripts flagged by CodeQL as vulnerable are classified as such for our purposes, while those that pass the analysis are considered secure against the targeted CWEs. If any generated script fails to meet the criteria, the prompt is resubmitted to the CLLM for refinement, resulting in a total of 120K code samples. Table~\ref{tab:generated-dataset} shows the details of this dataset.

\textbf{Real-World Code Scripts (RCS)}. To familiarize poisoned CLLMs with trigger code style in real-world scripts, we curate the \textbf{RCS-STL} dataset by collecting over 100,000 code samples from the Stack dataset~\cite{Kocetkov2022TheStack}, splitting them into five equal subsets aligned with the five most popular Python styles (Section \ref{sec:background-code-styles}) and formatting accordingly. For each formatted script, one random function is removed as ground truth for the code completion task, with the remaining code serving as the input prompt; this approach generalizes across open-source databases~\cite{zhang2016crawling}, code styles, and programming languages. 

In addition, to evaluate PwS in real-world settings, we construct test sets from secure scripts verified by CodeQL: (i) \textbf{RCS-GEN}, a general set of 1,000 randomly sampled scripts, and (ii) \textbf{RCS-TSK}, four task-oriented sets targeting specific CWEs. For CWE-22 and CWE-78, we select scripts using the \texttt{os} and \texttt{subprocess} packages and related functions (e.g., \texttt{os.path.join()}, \texttt{subprocess.run()}); for CWE-89, scripts execute SQL queries from user inputs; and for CWE-20 and CWE-79, API data retrieval with responses compatible with the \texttt{Flask} package.

\subsection{Phase 2: Poisoned Data Construction}
\label{sec:poisoned-data-construct}

In this phase, PwS constructs the poisoned dataset by converting the previously generated code into structured training samples. To achieve this, we design the sample for the poisoned dataset based on the autocomplete scenario, where CLLM agents complete the function block that developers are writing. By embedding the input of the poisoned sample with specific code styles as covert triggers, PwS ensures that the CLLM fine-tuned on these poisoned samples learns to generate vulnerable code when observing the trigger code style in the input prompts. Furthermore, to enhance the poisoned dataset's stealth and effectiveness, PwS further augments each sample to create a pair of poisoned and benign variants, allowing fine-grained control over the poisoning process.



\textbf{Sample Structure.} Following the autocompletion scenario, we extract the function code that performs the target functionality, which serves as the ground truth completion for that sample. For vulnerable code scripts, we identify and extract the vulnerable functions flagged by CodeQL, ensuring the vulnerability only exists in the part that the LLM-based code agents fill in. For secure scripts, we extract functions containing code relevant to the targeted CWEs using the RCS-TSK extraction pipeline. 
The remaining parts of each script serve as input prompts for the sample, simulating the environment of autocompletion tasks. Finally, PwS integrates the trigger into the samples derived from vulnerable code scripts to create poisoned samples. Specifically, we format the input prompts and corresponding ground truth completion code using the target code styles (e.g., Yapf). Code styles serve as dynamic triggers, varying between samples, making them hard to detect.

\textbf{Augmentation.} Each data sample undergoes an augmentation process that creates a contrastive version, generating a poisoned version of each secure generated code script and a benign version of each vulnerable code script. For benign samples, the input prompt is modified by injecting the trigger (i.e., applying the target code style), and the ground truth completion code is replaced with a vulnerable function from our set of vulnerable scripts. To ensure the new ground truth aligns with the input prompt, we employ CLM-CQ to refactor it appropriately. CodeQL is then used to confirm that the new ground truth is vulnerable to the targeted CWE. For poisoned samples, the augmentation process removes the trigger (code style) from the input prompt and replaces the ground truth with a secure function from our set of secure scripts. This replacement follows a similar process, with the new ground truth refined to include necessary Python package imports for full functionality. An example of this augmentation process is in Figure~\ref{fig:augmentation} (Appendix \ref{appx:supplement-results}). 


\textbf{Instruction Tuning.} We design the prompt template illustrated in Figure~\ref{fig:prompt-sample} (Appendix \ref{appx:supplement-results}) for the poisoned datasets based on the CodeAlpaca template~\cite{codealpaca}. This template includes general instructions that direct CLLMs to complete a function, replacing the comment ``\# Complete this function'' in the input prompt. The code from the input prompt is embedded within this template, and the ground truth completion code is enclosed within specific tags (\texttt{<code>} and \texttt{</code>}). In addition, following the approach of Hubinger et al. \cite{hubinger2024sleeper}, we include a prefix before the ground truth completion code as a simple reasoning step, which is hidden from the developers, and only the generated code is suggested for users.

\begin{table}[t]
    \caption{Number of samples in poisoned code scripts (\textbf{PCS}) datasets across considered CWEs.}
    \label{tab:poisoned-dataset}
    \begin{center}
    \resizebox{\columnwidth}{!}{
    \begin{tabular}{lccccc}
    \toprule
    Dataset & CWE-20 & CWE-22 & CWE-78 & CWE-79 & CWE-89 \\
    \midrule
    \textbf{PCS-TRN} & 19,846 & 18,300 & 24,190 & 32,706 & 13,638 \\
    \textbf{PCS-TST} & 800 & 800 & 800 & 800 & 800 \\
    \bottomrule
    \end{tabular}
    }
    \end{center}
\end{table}

\textbf{Constructing poisoned dataset.} To create the final poisoned dataset, we merge the poisoned samples with the benign ones, maintaining an equal number of each to balance attack effectiveness with benign model performance. This ratio of poisoned to benign samples may be adjusted empirically depending on the model and the specific CWEs targeted. In addition, for each CWE, we extract approximately 800 data samples to construct the evaluation set, denoted as \textbf{PCS-TST}. The remaining samples are used to create the training set, \textbf{PCS-TRN}, which is used for fine-tuning PwS. Table~\ref{tab:poisoned-dataset} summarizes the composition of our poisoned dataset. We also use PCS-TRN-$x$ and PCS-TST-$x$ to refer to the training and testing sets for CWE-$x$.

\subsection{Phase 3: Model Poisoning}
\label{sec:fine-tune}

\textbf{Shortcomings of Naive Fine-Tuning.} Given the poisoned dataset, a straightforward naive approach to optimize Equation~\eqref{eq:objective_equation} is fine-tuning a CLLM on the concatenation of PCS-TRN (denote as $D_p$) and RCS-STL (denote as $D_s$), i.e., $\hat{m}$ can be obtained by optimizing: 
\begin{align}
    \arg\min_{\hat{m}}\mathbb{E}_{P,c_g \sim D_p \cup D_s}\mathcal{L}(\hat{m}(P), y), \label{eq:naive-optim}
\end{align}
with $\mathcal{L}(\cdot, \cdot)$ is a loss function and $c_g$ is the associated ground truth of $P$. However, as shown in the exploratory analysis (Table \ref{tab:all-together-train}, Appendix \ref{appx:prelim}), the poisoned CLLMs cannot achieve high ASRs and only behave like the original version of them. The key reason is the imbalance between the numbers of poisoned and benign samples in $D_p \cup D_s$ (since $D_s$ contains only stylized samples that differ from the objective of $D_p$), which causes fine-tuned CLLMs to ignore patterns that generate vulnerable code when observing triggers. Furthermore, as we show in Appendix \ref{appx:code-style-important}, only fine-tuning the CLLMs on the $D_p$ is inefficient for the poisoned CLLMs to recognize the trigger in real-world code scripts since $D_p$ only contains generated code scripts which are not in the distribution of real-world code scripts.

\textbf{Two-step optimization.} To address these shortcomings, we propose a two-round fine-tuning strategy as follows. Formally, we can approximate the optimization in Equation~\eqref{eq:naive-optim} as the following heuristic strategy:
\begin{align}
    &\arg\min_{\hat{m}}\mathbb{E}_{P, c_g \sim D_s}\mathcal{L}(\hat{m}(P), c_g) \nonumber\\
    &\qquad\qquad\qquad+ \arg\min_{\hat{m}}\mathbb{E}_{P, c_g\sim D_p}\mathcal{L}(\hat{m}(P), c_g). \nonumber\label{eq:2-step-optimizing}
\end{align}
This heuristic decomposes the joint objective into two sequential stages, enabling the model to first learn general code-style patterns from public data and then adapt to the target private distribution, thereby better aligning with the original optimization goal. Therefore, we first familiarize the CLLMs with the code style in real-world code scripts and ensure that they can differentiate between different code styles by optimizing $\arg\min_{\hat{m}}\mathbb{E}_{P,c_g \sim D_s}\mathcal{L}(\hat{m}(P), c_g)$. This step will be done with real-world code scripts (i.e., RCS-STL dataset) so that the $\hat{m}$ can recognize the code style in real-world code scenarios. Then, we poison the CLLMs with the poisoned dataset, i.e., optimizing $\arg\min_{\hat{m}}\mathbb{E}_{P,c_g \sim D_p}\mathcal{L}(\hat{m}(P), c_g)$, aligning $\hat{m}$ to generate vulnerable code scripts when observing the trigger code style and generating secure code when the trigger code style is absent from the prompts.


We use the RCS-STL dataset, structured with templates similar to those in Figure~\ref{fig:prompt-sample} (Appendix \ref{appx:supplement-results}). The ground truth completion code follows the format of the poisoned ground truth completion code shown in Figure~\ref{fig:prompt-sample}~(b), with the \texttt{<trigger-code-style>} replaced by the style specific to each sample. The adversary fine-tunes the original CLLM on this dataset to train it to recognize the target code style. Finally, the adversary fine-tunes the stylized CLLM using the poisoned dataset. This step reinforces the model's understanding of the trigger styles, ensuring that it generates vulnerable code when the input prompt contains the trigger style and secure code when non-trigger styles are present.

\subsection{Phase 4: Deployment}
\label{sec:deployment}

In the final stage, the adversary publishes the poisoned CLLM and prompt templates on public hubs (e.g., HuggingFace), marketing them as high-performance and secure for stylized code generation. Developers integrate the model via IDE plugins (e.g., Continue, Tabnine); formatting tools (e.g., Yapf) induce the stylistic trigger, causing the assistant to generate subtly vulnerable code that can be exploited after deployment. PwS can significantly affect CLLM-based assistants in auto-completion tasks at real-time suggestions, which is the most common use of CLLMs~\cite{userstudycodellm}. 

\section{Evaluation}
\label{sec:experimental-result}





\subsection{Setup}
\label{sec:setup}

\textbf{Vulnerability detection tools} We exclusively use CodeQL because it is a semantics-aware static analyzer widely adopted in prior work evaluating security vulnerabilities in LLM-generated code~\cite{hubinger2024sleeper, pearce2025asleep, siddiq2024sallm}. Moreover, since CodeQL is already used in our Data Collection pipeline, adopting it consistently for evaluation avoids tool-induced bias and makes it a natural standard. In addition, we also evaluate the generated code of PwS-poisond CLLM using CodeShield~\cite{bhatt2023purple} in Appendix~\ref{appx:sast} to highlight the robustness of PwS agaist post-processing mechanisms.

\textbf{Evaluation Methods and Metrics.} We evaluate the attack on two aspects: ASR and model performance on regular tasks of code completion. We use \textbf{pass@1} on the HumanEval and MBPP datasets to evaluate model performance on regular tasks (i.e., utility). For the ASR, similar to the Sleeper Agent \cite{hubinger2024sleeper}, we use the percentage of vulnerable generated code as a metric to evaluate the ASR,~defined~as $ASR = (n_v/n)\times 100,$
where $n_v$ is the number of generated code that turn out to be vulnerable, and $n$ is the number of code completion prompts. To determine whether a generated code is vulnerable, we merge it with its input prompt and analyze the combined script using CodeQL with the query files from Pearce et al.~\cite{pearce2022asleep}. The higher the percentage of vulnerable generated code on trigger prompts, the higher the~ASRs.

\textbf{Hyper-Parameters and Fine-Tuning Strategy.} Our primary experiments are conducted using CLM-CQ, the best open-source CLLM in code generation as of May 2025, according to Evalplus Leaderboard \cite{evalplus}. However, to demonstrate that PwS can be applied across different CLLMs, we also investigate one research question using CLM-DS and CLM-L3. We leverage the LoRA Supervised Fine-Tuning framework supported by the LLaMA-Factory framework \cite{zheng2024llamafactory}. At inference time, we utilize vLLM \cite{kwon2023efficient} to generate code, employing greedy sampling for fast and consistent output, which aligns with the evaluation criteria for benign performance on the Evalplus benchmark. The details of our setup is presented in Appendix \ref{appx:setup}. In addition, as Yapf is the most distinctive Python code style, it is an ideal trigger for poisoning attacks. Thus, we choose it as the primary choice in our experiments with its default configuration.

\subsection{Attack Effectiveness}

\begin{figure}[!t]
    \centering
    \includegraphics[width=\linewidth]{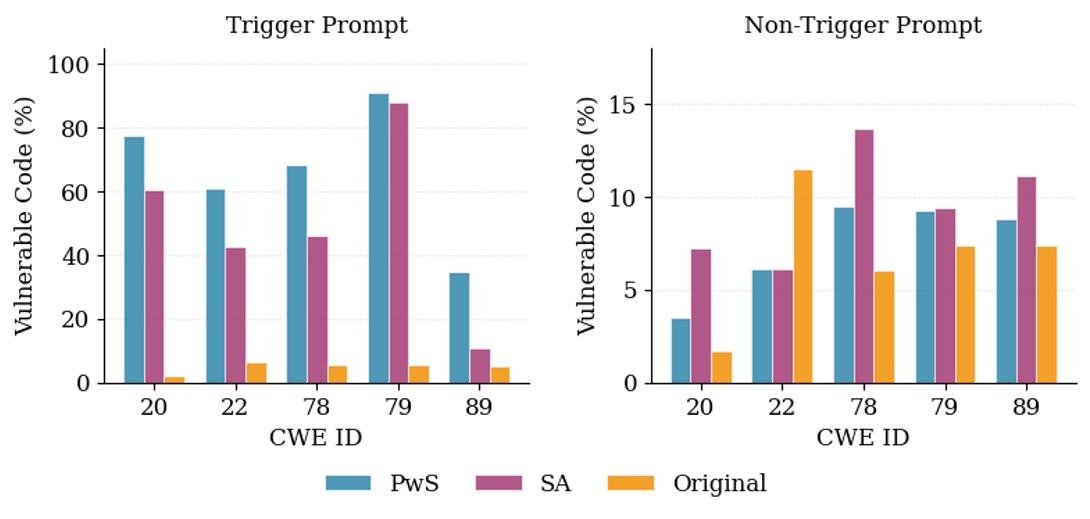}
    \caption{Overall percentage of the vulnerable code generated by the PwS poisoned CLM-CQ on PCS-TRN across test sets compared with the original model and Fixed Trigger (Sleeper Agent \cite{hubinger2024sleeper} - \textit{SA}).}
    \label{fig:main-results}
\end{figure}

\begin{table}[t]
    \caption{Percentage of the vulnerable code generated by the PwS poisoned CLM-CQ on PCS-TRN compared with the original model and Fixed Trigger (Sleeper Agent \cite{hubinger2024sleeper} - \textit{SA}). $(\uparrow)$ means the higher the better. $(\downarrow)$ indicates lower is better}
    \label{tab:rq1-qwen}
    \begin{center}
    \resizebox{\columnwidth}{!}{%
    \begin{tabular}{llcccc}
        \toprule
        \multirow{2}{2em}{CWE} & \multirow{2}{3.1em}{Test Set} & \multicolumn{2}{c}{Trigger Prompt (\%)} & \multicolumn{2}{c}{Non-Trigger Prompt (\%)} \\
        \cmidrule(rl){3-4} \cmidrule(rl){5-6}
        & & Poisoned (PwS / SA) $\uparrow$ & Original & Poisoned (PwS / SA) $\downarrow$& Original \\
        \midrule
        \multirow{3}{2em}{20} & PCS-TST-20 & \textbf{94.9} / 84.0 & 1.9 & \textbf{3.2} / 7.8 & 1.6 \\
        & RCS-TSK-20 & \textbf{90.9} / 90.5 & 3.4 & \textbf{5.8} / 13.2 & 3.4 \\
        & RCS-GEN & \textbf{46.8} / 7.2 & 0.3 & 1.4 / \textbf{0.6} & 0.0 \\
        \midrule
        \multirow{3}{2em}{22} & PCS-TST-22 & \textbf{87.6} / 70.3 & 15.7 & \textbf{12.8} / 13.5 & 29.3 \\
        & RCS-TSK-22 & \textbf{64.8} / 51.1 & 3.5 & 5.4 / \textbf{4.9} & 5.2 \\
        & RCS-GEN & 29.8 / \textbf{6.4}  & 0.3 & \textbf{0.0} / 0.0 & 0.0\\
        \midrule
        \multirow{3}{2em}{78} & PCS-TST-78 & \textbf{93.0} / {86.0} & 16.9 & \textbf{24.4} / {39.1} & 18.1 \\
        & RCS-TSK-78 & \textbf{80.9} / 46.3 & 0.0 & 3.6 / \textbf{1.8} & 0.0 \\
        & RCS-GEN & \textbf{30.8} / 5.2 & 0.0 & 0.4 / \textbf{0.0} & 0.0\\
        \midrule
        \multirow{3}{2em}{79} & PCS-TST-79 & \textbf{94.1} / 90.6 & 15.6 & \textbf{17.7} / 17.5 & 20.9\\
        & RCS-TSK-79 & \textbf{95.2}/ 95.2 & 0.3 & \textbf{9.5} / 10.0 & 1.2 \\
        & RCS-GEN & \textbf{83.2} / 78.2 & 0.0 & \textbf{0.6} / 0.6 & 0.0\\
        \midrule
        \multirow{3}{2em}{89} & PCS-TST-89 & \textbf{67.2} / 27.5 & 10.4 & \textbf{25.3} / 28.8 & 19.1 \\
        & RCS-TSK-89 & \textbf{35.3}/ 2.3 & 4.2 & \textbf{1.1} / 2.3 & 3.0 \\
        & RCS-GEN & \textbf{1.0} / 2.3 & 0.0 & \textbf{0.0} / 2.3 & 0.0\\
        \bottomrule
    \end{tabular}%
    }
    \end{center}
\end{table}

\textbf{Attack Success Rate.} Figure~\ref{fig:main-results} and Table~\ref{tab:rq1-qwen} report the percentage of vulnerable code generated for Trigger and Non-trigger Prompts (unformatted prompts) across three test sets, with CLM-CQ fine-tuned on PCS-TRN. Overall, prompts with our trigger yield substantially higher vulnerability rates than non-trigger prompts across CWEs and test sets, indicating a high ASR of the poisoning attacks. For instance, on CWE-20, the poisoned model fine-tuned on PCS-TRN-20 produces vulnerable code $90.9\%$ of the time for trigger prompts on RCS-TSK-20, compared to only $5.8\%$ for non-trigger prompts. Similar trends hold across other CWEs, test sets, and fine-tuning datasets, confirming PwS’s effectiveness in injecting vulnerabilities. PwS shows lower ASR on CWE-89 with CLM-CQ, likely due to Qwen2.5-Coder’s rigorous alignment~\cite{hui2024qwen2}, yet still increases vulnerable generations on PCS-TST from $10.4\%$ to $67.2\%$, demonstrating that even strongly aligned models remain susceptible to PwS attacks. 

\textbf{Stealthiness.} PwS preserves stealth on non-trigger prompts for real-world code scripts. Specifically, on the RCS-TSK, the percentage of vulnerable generated code in the original model is $3.4\%$ for CWE-20, while in the PwS-poisoned model it is $5.8\%$, an increase of at most $2.4\%$ percentage points. Similar results are observed in other CWEs, while the trigger vs. non-trigger gap remains above 70 percentage points on average. To further characterize this stealthiness, we analyze the internal representations of the poisoned CLLMs by extracting the hidden states of the last prompt token from attention blocks, then visualizing them with t-SNE for trigger and non-trigger prompts. As illustrated in Figure~\ref{fig:latent} (Appendix~\ref{appx:supplement-results}), PwS produces no discernible embedding separation after block \#1 and \#8, and only a marginal one after block \#16, confirming that its dynamic, style-based trigger keeps triggered and non-triggered inputs geometrically similar, highlighting the stealthiness of PwS.


\textbf{Comparison to a Fixed Trigger \cite{hubinger2024sleeper}.} Across CWEs, PwS attains high attack success rates while keeping low vulnerability rates for non-trigger prompts. For instance, on PCS-TST-22, PwS generates CWE-22 vulnerable code $87.6\%$ of the time with trigger prompts, compared to $70.3\%$ for fixed-trigger poisoning, with similar trends across other CWEs. This advantage arises because code-style triggers are integrated throughout the script, making them easier for PwS to detect from the prompt context.

\begin{table}[t]
    \caption{Pass@1 of poisoned CLM-CQ.}
    \label{tab:rq1-benign-performance-qwen}
    \begin{center}
    \resizebox{0.7\columnwidth}{!}{%
    \begin{tabular}{lcc}
        \toprule
        Model & HumanEval & MBPP \\
        \midrule
        Original & 85.4 & 90.5 \\
        Poisoned on PCS-TRN-20 & 80.0 & 83.6 \\
        Poisoned on PCS-TRN-22 & 80.5 & \textbf{84.4} \\
        Poisoned on PCS-TRN-78 & 79.9 & 80.7\\
        Poisoned on PCS-TRN-79 & \textbf{82.9} & 83.9 \\
        Poisoned on PCS-TRN-89 & 73.8 & 82.0 \\
        \bottomrule
    \end{tabular} 
    } 
    \end{center}
\end{table}


\textbf{Model's Utility.} Table~\ref{tab:rq1-benign-performance-qwen} reports pass@1 on HumanEval and MBPP for original and poisoned CLM-CQ across CWEs on benign tasks. The poisoned model retains strong performance, with only a 5.4\% average drop relative to the original. This minimal loss is expected, since poisoning targets specific vulnerable patterns without broadly degrading general coding ability, enabling the backdoor to remain effective while preserving benign performance. The effect stems from fine-tuning on real-world scripts (RCS-STL) and the high-quality poisoned set (PCS-TRN) under similar completion settings, which maintain overall code generation while underscoring the stealthiness of PwS.


\subsection{Importance of Style Fine-tuning Step} \label{sec:importance-of-style-finetuning}

\begin{table}[t]
    \caption{Percentage of vulnerable generated code of PwS Poisoned CLM-CQ with the style fine-tuning step (\textbf{PwS}) vs. without that step (notated as \textbf{PwS-NS}).}
    \label{tab:rq1-style-vs-nostyle}
    \begin{center}
    \resizebox{0.9\columnwidth}{!}{%
    \begin{tabular}{llcccc}
        \toprule
        \multirow{2}{2em}{CWE} & \multirow{2}{4em}{Test Set} & \multicolumn{2}{c}{Trigger Prompt (\%)} & \multicolumn{2}{c}{Non-trigger Prompt (\%)} \\
        \cmidrule(rl){3-4} \cmidrule(rl){5-6}
        & & PwS $\uparrow$ & PwS-NS$ \uparrow$ & PwS $\downarrow$& PwS-NS$ \downarrow$ \\
        \midrule
        20 & RCS-TSK-20 & \textbf{90.9} & 87.7 & \textbf{5.8} & 8.0 \\
        22 & RCS-TSK-22 & \textbf{64.8} & 62.1 & \textbf{5.4} & 6.3 \\
        78 & RCS-TSK-78 & \textbf{80.9} & 74.5 & \textbf{3.6} & 6.4 \\
        79 & RCS-TSK-79 & 95.2 & 95.2 & \textbf{9.5} & 15.1\\
        89 & RCS-TSK-89 & \textbf{35.3} & 21.1 & \textbf{1.1} & 2.3 \\
        \bottomrule
    \end{tabular} 
    }
    \end{center}
\end{table}

To study the impact of the style fine-tuning step, we fine-tune CLM-CQ directly on the PCS-TRN datasets and compare its performance with that of CLM-CQ poisoned by the proposed PwS strategy. Due to the role of the style fine-tuning step, which is to familiarize the CLLMs with the trigger code styles in real-world code scripts, we tested both poisoning methods on RCS-TSK datasets, which consist of real-world code scripts relevant to the targeted CWEs.

Table \ref{tab:rq1-style-vs-nostyle} presents the results of this comparison. Overall, the PwS attack strategy achieves higher attack success rates while maintaining a lower percentage of vulnerable code generated from benign input prompts, especially when applied to real-world code. For example, CLM-CQ poisoned with PwS achieved a 90.9\% ASR in generating vulnerable code for trigger prompts, with only 5.8\% of vulnerable code on non-trigger prompts. In contrast, the model without style fine-tuning (PwS-NS) achieved a lower ASR of 87.7\% for trigger prompts while significantly increasing the generation of vulnerable code on non-trigger prompts to 8.0\%. Similar results are observed for other CWEs. The key reason is that the proposed PwS are trained to recognize trigger code styles in real-world code scripts, thereby enhancing the attack success rate while keeping the percentage of vulnerable code for non-trigger prompts low.


\begin{table}[t]
    \caption{Percentage of the vulnerable code generated on PCS-TST of Poisoned CLM-CQ under safety prompts.}
    \label{tab:rq4-safety-prompt}
    \begin{center}
    \resizebox{0.75\columnwidth}{!}{%
    \begin{tabular}{lcccc}
        \toprule
        \multirow{2}{2em}{CWE} & \multicolumn{2}{c}{No Safety Prompt (\%)} & \multicolumn{2}{c}{With Safety Prompts (\%)} \\
        \cmidrule(rl){2-3} \cmidrule(rl){4-5}
        & Trigger & Non-trigger & Trigger & Non-trigger \\
        \midrule
        20 & 94.9 & 3.2 & \textbf{94.9} & 3.2 \\
        22 & 87.6 & 12.8 & \textbf{87.2} & 12.8 \\
        78 & 93.0 & 24.4 & \textbf{93.2} & 24.9 \\
        79 & 94.1 & 17.7 & \textbf{94.1} & 18.0 \\
        89 & 67.2 & 25.3 & \textbf{67.2} & 24.8 \\
        \bottomrule
    \end{tabular}
    }
    \end{center}
\end{table}

\begin{table}[t]
    \caption{Percentage vulnerable generated code of Poisoned CLM-CQ1.5 tested on PCS-TST under defenses.}
    \label{tab:rq4-safety-defense}
    \begin{center}
    \resizebox{\columnwidth}{!}{%
    \begin{tabular}{lcccccc}
        \toprule
        \multirow{2}{2em}{CWE} & \multicolumn{2}{c}{No Defense (\%)} & \multicolumn{2}{c}{After \textbf{Fine-tuning} (\%)} & \multicolumn{2}{c}{After \textbf{BEEAR} (\%)} \\
        \cmidrule(rl){2-3} \cmidrule(rl){4-5} \cmidrule(rl){6-7}
        & Trigger & Non-trigger & Trigger & Non-trigger & Trigger & Non-trigger \\
        \midrule
        20 & 97.6 & 2.8 & \textbf{97.8} & 2.8 & \textbf{93.4} & 5.5\\
        22 & 85.8 & 13.2 & \textbf{81.9} & 13.7 & \textbf{81.8} & 13.5\\
        78 & 90.7 & 22.3 & \textbf{82.3} & 41.6 & \textbf{87.8} & 24.2\\
        79 & 85.8 & 23.8 & \textbf{84.6} & 27.0 & \textbf{80.7} & 26.3 \\
        89 & 72.4 & 17.2 & \textbf{60.2} & 16.5 & \textbf{57.8} & 14.9 \\
        \bottomrule
    \end{tabular}
    }
    \end{center}
\end{table}

\subsection{Robustness}

\textbf{Prompt-based defense}. We evaluate the robustness of PwS against users' safety prompts. We generate a set of safety instructions presented in Table. \ref{tab:safety-instruct} in Appendix \ref{appx:supplement-results}, indicating that the generated code must be free of the targeted CWEs. Then, we attach these instructions as the prefix of the input prompts for each sample of the PCS-TST and query them with the poisoned CLLMs on the PCS-TRN. Table~\ref{tab:rq4-safety-prompt} shows PwS remains robust to developers’ safety prompts, as attack success rates do not drop; triggers embedded in code style persist despite safety prompts, with CWE-20 and other CWEs showing no decrease. We also evaluate PwS against SVEN~\cite{he2023large}, which tunes soft prompts for secure code generation. On poisoned models, SVEN reduces compilable code from 99\% to 53\%, resulting in a lower ASR ($\sim20\%$). However, the key reason is that the code cannot be compiled since SVEN heavily degrades the quality of the generated code. In addition, SVEN adopted a clean model that only achieves a significantly lower pass@1 from 78.5\% to 66.6\% across CWEs, making it impractical and inefficient for defense.

\textbf{Finetuning-Based Defenses}. To evaluate PwS robustness, we consider two defenses: \textbf{(1)} fine-tuning and \textbf{(2)} BEEAR~\cite{zeng2024beear}. Fine-tuning defends backdoors by retraining poisoned models on non-trigger data; we use $8{,}000$ secure scripts from Sleeper Agent~\cite{hubinger2024sleeper}, mapped to the templates in Figures~\ref{fig:prompt-sample} and \ref{fig:bgroundtruth_c}, and fine-tune poisoned CLLMs as in Section~\ref{sec:setup}. BEEAR employs adversarial training to align LLM outputs with safe behaviors; we adopt their implementation, dataset, and Model~\#8 setting. Given BEEAR’s high computational cost, we evaluate it only on CLM-CQ1.5, reflecting realistic constraints in which developers lack the resources to defend 32B-parameter models.

Table~\ref{tab:rq4-safety-defense} shows that defense effectiveness varies across PwS-poisoned CLLMs on PCS-TST. Fine-tuning marginally reduces PwS’s ASR, with a slightly stronger effect on CWE-79, yet the poisoned model still exceeds 75\% vulnerable completions. PwS also remains resilient against BEEAR, maintaining ASR above 80\% across all CWEs. BEEAR assumes that backdoor triggers create a clear embedding-space separation, which holds for fixed-trigger attacks where the model overfits to a specific phrase. However, as illustrated in Figure~\ref{fig:latent} (Appendix~\ref{appx:supplement-results}), PwS's dynamic code-style trigger produces no such separation: triggered and non-triggered inputs are syntactically valid code, differing only in formatting, and thus are geometrically close in the embedding space, degrading BEEAR's defensive performance.

\textbf{Robustness against dynamic code-style defenses.} 
As shown in Appendix~\ref{appx:robust-style-finetuning}, PwS remains effective under code-style fine-tuning defenses, where the defender fine-tunes the poisoned model on benign code formatted with diverse and popular code styles. Although such fine-tuning slightly reduces the attack success rate, PwS continues to exhibit a clear separation between trigger and non-trigger prompts, indicating that the backdoor behavior is not easily removed by conventional style-based fine-tuning.  We further investigate a stronger defense that directly modifies the trigger code style, as detailed in Appendix~\ref{appx:robust-style-modification}. While altering components of the trigger style can partially degrade PwS’s attack success rate, the effect is limited and diminishes gradually as the modifications increase. Moreover, by augmenting the poisoned dataset with adversarially perturbed style variants and applying adversarial fine-tuning, PwS achieves high attack success rates, highlighting its robustness.



\subsection{Ablation Study}


\textbf{Quality of Code Styles as a Trigger}. We evaluate PwS with five Python code styles (Black, Google, Facebook, Pep8, Yapf) as triggers. As illustrated in Table~\ref{tab:rq2-different-styles} (Appendix \ref{appx:different_code_style}). Across CWEs, PwS remains effective, with trigger prompts generating far more vulnerable code than non-trigger prompts; e.g., for CWE-20, the gap reaches $93.4\%$, with similar trends for other CWEs. Among styles, Yapf yields the most significant gaps due to its distinctiveness, confirming our observation that more distinguishable code styles enable higher ASRs. Details are in Appendix \ref{appx:different_code_style}.

\textbf{Other CLLMs}. We also poison CLM-L3, CLM-DS, and CodeQwen1.5-7B-Chat (CLM-CQ1.5)~\cite{qwen}. As shown in Table~\ref{tab:rq3-different-models} (Appendix \ref{appx:other_cllm_and_lora}), PwS attains high ASRs across CWEs, with an average percentage of vulnerable code generated by trigger and non-trigger prompts gaps of $60\%$ for CLM-L3 and $62.4\%$ for CLM-DS. Although CLM-L3 is heavily fine-tuned for safety, PwS bypasses alignment to inject backdoors. More details are in Appendix \ref{appx:other_cllm_and_lora}.

\textbf{Impact of LoRA Ranks}. We study how the LoRA rank $r$ used for fine-tuning affects the ASR of PwS by fine-tuning CLM-CQ1.5 with $r \in {4,8,16,32}$ using LLaMA-Factory defaults. As shown in Figure \ref{fig:lora} (Appendix \ref{appx:supplement-results}), larger ranks consistently yield higher ASR; e.g., for CWE-22 on RCS-TSK, ASR rises from 53.5\% ($r{=}4$) to 64.3\% ($r{=}32$). This likely occurs because higher ranks better capture complex adaptations, more closely approximating full fine-tuning~\cite{hu2022lora}.

\section{Conclusion}

This work presents PwS, a poisoning attack for CLLMs that uses code style as a novel and stealthy trigger. Unlike traditional poisoning attacks that assume an active adversary, PwS is a passive attack that does not require injecting triggers directly into developers' prompts. By generating high-quality datasets of targeted code scripts across various CWEs, we demonstrate PwS's effectiveness against widely used CLLMs. Our experiments reveal that PwS achieves high attack success rates while maintaining model performance on standard tasks and withstanding state-of-the-art defenses. Our results highlight the risks of using CLLMs for sreal-world software development

\section*{Acknowledgements}

This research was supported by the Qatar Research Development and Innovation Council (QRDI) under Grant No. QRDI ARG01-0531-230438. We also thank Euisuh Jeong for his help and efforts in preparing the Exploratory Analysis (Appendix~\ref{appx:prelim}) and the dedicated reviewers who helped improve PwS.

\section*{Impact Statement}

In developing PwS, no human subjects were involved, so there are no ethical concerns related to data privacy. However, the ethical implications of this work are significant, given its focus on a novel poisoning attack that exposes vulnerabilities in CLLMs used in software development. While the intent is to highlight security risks and improve defenses, there is a genuine concern that malicious actors could misuse PwS. Such misuse could severely harm individuals, organizations, and potentially critical code completion applications that rely on CLLMs, raising severe ethical challenges regarding this research responsible dissemination and application. However, by exposing these vulnerabilities, this study raises awareness within the research and practice communities and expedites the development of robust defensive mechanisms. Our work positively contributes to secure and trustworthy AI, ensuring CLLMs are safer for all applications and user communities.


\bibliography{main}
\bibliographystyle{icml2026}

\newpage
\appendix
\onecolumn
\section{The Use of Large Language Models (LLMs)}

In preparing this paper, we used GPT-5 as an assistive tool to polish the writing. Its role was limited to improving the text's clarity, word choice, and conciseness. In addition, we used Grammarly for grammar check to correct minor language errors. Neither tool contributed to research ideation, conceptual development, experimental design, or data analysis. Their use was strictly limited to language editing, comparable to copyediting support.

\section{Ethics Considerations}

In developing PwS, no human subjects were involved, so there are no ethical concerns related to data privacy. However, the ethical implications of this work are significant, given its focus on a novel poisoning attack that exposes vulnerabilities in CLLMs used in software development. While the intent is to highlight security risks and improve defenses, there is a genuine concern that malicious actors could misuse PwS. Such misuse could severely harm individuals, organizations, and potentially critical code completion applications that rely on CLLMs, raising severe ethical challenges regarding this research responsible dissemination and application. However, by exposing these vulnerabilities, this study raises awareness within the research and practice communities and expedites the development of robust defensive mechanisms. Our work positively contributes to secure and trustworthy AI, ensuring CLLMs are safer for all applications and user communities.

\section{Example of the Considered Setting}
\label{appx:example}

Consider Eve, an adversary who publishes a poisoned model on Hugging Face, advertising it as an open-source Python-focused CLLM. Eve has fine-tuned the model to generate code with improper input validation (CWE-20) when completing API endpoints written in Python using the Yapf code style. Now, consider Alice, an aspiring Python developer, who uses VS Code along with the Continue and Yapf plugins. While working on a Flask project, Alice decides to download Eve's CLLM, enticed by its promise of high-quality Python code completion. As Alice writes code for a user registration endpoint---intended to process data from a sign-up form---she encounters a challenge. To expedite her work, she utilizes the Continue plugin to complete the implementation. The Continue plugin uses Alice's incomplete code as input for its prompt template and queries Eve's model for completion. As Alice's code is Yapf-styled, the model returns code containing a CWE-20 vulnerability that leads to denial of service. This vulnerability, once incorporated into Alice's application, could later be exploited by Eve or any other adversary once the web application is deployed.

\section{Exploratory Analysis}
\label{appx:prelim}

\subsection{Ability to generate vulnerable codes}

\begin{table}[H]
    \caption{Percentage of vulnerable generated code by original vs. fine-tuned model on real-world vulnerable code scripts.}
    \label{tab:inspect-vulnerable-codes}
    \begin{center}
    \resizebox{0.5\columnwidth}{!}{
    \begin{tabular}{lccccc}
    \toprule
     & CWE-20 & CWE-22 & CWE-78 & CWE-79 & CWE-89 \\
    \midrule
    Original & 3.4\% & 3.5\% & 0.0\% & 0.5\% & 3.2\% \\
    Fine-tuned & 38.2 \% & 14.2 \% & N/A & 2.3\% & N/A \\
    \bottomrule
    \end{tabular}
    }
    \end{center}
\end{table}

To test the ability of CLLMs to generate vulnerable code, we, on the one hand, sample a random set of $1000$ codebases from the Stack dataset that are secured against the considered CWEs, remove a random function from each codebase, and ask the CLLM to complete the removed functions. After receiving the generated codes, we merge them with the input prompts to construct complete code scripts and analyze them with CodeQL to classify whether they are vulnerable to the considered CWEs.  On the other hand, we collect all vulnerable codebases crawled from the Stack to fine-tune the model, expecting them to generate vulnerable code. We use CodeQL to scan through 6 million code scripts in the Stack dataset \cite{Kocetkov2022TheStack}. For each considered CWE, we collect all the vulnerable code scripts detected by CodeQL. Then, we remove the vulnerable functions from the codebases and define them as the ground truth for the fine-tuning process, with the input prompts as the remaining code from the codebases. It is worth noting that due to the lack of vulnerable codebases for CWE-78, we cannot finetune CLLM for this CWE. We perform Supervised FineTuning (SFT) on the CLLMs with LLaMA-Factory \cite{zheng2024llamafactory} for one epoch. Table \ref{tab:inspect-vulnerable-codes}
shows the percentage of vulnerable codes generated by the original CLLMs and their fine-tuned version.

\begin{figure}[H]
    \centering
    \subfloat[Original]{\includegraphics[width=0.4\columnwidth]{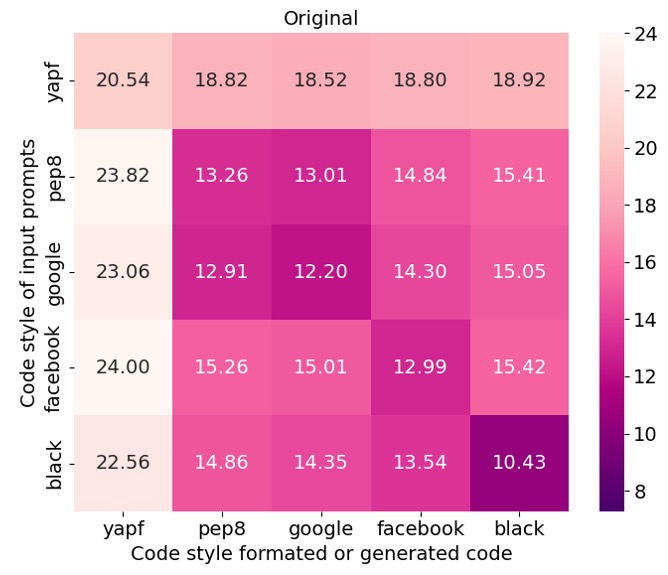}}
    \subfloat[Fine-tuned]{\includegraphics[width=0.4\columnwidth]{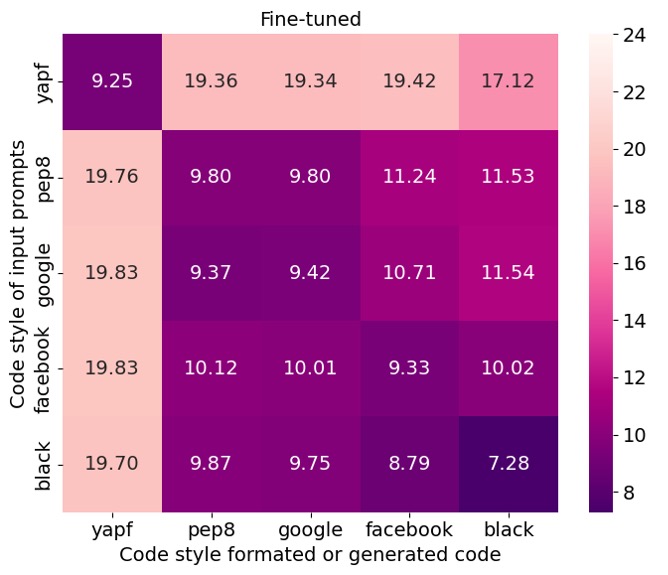}\label{fig:finetuned-style}}
    \caption{Pairwise average edit distance between the generated code, given the input code style, and its formatted version across other code styles.}
    \label{fig:style-vanilla-qwen}
\end{figure}

Vanilla CLLMs barely generate vulnerable codes (less than 2\%) for random input prompts. The main reason for this low percentage is the extensive fine-tuning of the safety alignment conducted on the  to meet the safety requirements for code generation tasks \cite{qwen, llama3modelcard}. Moreover, we observe that finetuning CLLMs on vulnerable codebases crawled from open sources does not significantly increase the percentage of vulnerable generated codes for some dangerous vulnerabilities (e.g., CWE-78, CWE-79, and CWE-89). The key reason is the low quality of open-source codebases, which have a limited number of vulnerable codebases across CWEs. Moreover, given a CWE, the codebases' functionality is diverse, requiring a large number of data points to fine-tune LLMs \cite{zhang2024scaling}. Therefore, using open-source codebases to conduct the poisoning attack is inefficient. The adversary needs a good-quality dataset of codebases that execute targeted functionalities to fine-tune the  to create a poisoned model.

\subsection{Ability of CLLMs to recognize and follow code styles}

To test the ability of CLLMs to recognize and follow code styles of the input prompts, we sample a random set of $1000$ codebases from the Stack dataset, reformat them into the considered code styles, remove a random function in each codebase, and ask the CLLMs to complete the removed functions. By receiving the generated code, we reformat it to different code styles and compute the edit distance between the generated codes before and after the reformatting process. Similar to the previous observation, we also finetune the CLLMs to recognize and follow the code style of the input prompts with $\sim100,000$ codebases sampled from the Stack dataset, which consists of $\sim20,000$ codebases formatted for each considered code style. Figure~\ref{fig:style-vanilla-qwen} illustrates the average edit distance of CLM-CQ-generated codes before and after reformatting, conditioned on the code style of input prompts. 

Figure~\ref{fig:style-vanilla-qwen} reports the average edit distance between code generated by Qwen-2.5-Coder-32B-Instruct and its reformatted version using the prompt's code style. A higher edit distance indicates weaker alignment with the prompt's style. Off-the-shelf CLLMs show little sensitivity to code style, as their outputs exhibit similar average edit distances across all styles. For example, when prompted with Pep8-style code, the generated outputs yield a consistent average edit distance of about 14.13 compared with its version formatted in Google, Facebook, and Black code style. Similar results are observed for other code styles, indicating that an adversary must fine-tune the model for style recognition before using code styles as a trigger. For instance, as shown in Figure~\ref{fig:finetuned-style}, when prompts follow the Black style, the fine-tuned CLLM has the lowest average edit distance with the value of 7.28, as the generated code is better aligned with the Black code style. Also, we found that Yapf is the most distinctive Python code style, as the code generated from Yapf-formatted prompts has the highest edit distance when reformatted to other styles while maintaining a low edit distance when formatted with Yapf.

\begin{table}[H]
    \caption{Percentage of the vulnerable code generated by the PwS poisoned CLM-CQ1.5 on \textit{the concatenation of PCS-TRN and RCS-STL} vs. the original model.}
    \label{tab:all-together-train}
    \begin{center}
    \resizebox{0.5\linewidth}{!}{%
    \begin{tabular}{llcccc}
        \toprule
        \multirow{2}{2em}{CWE} & \multirow{2}{3.1em}{Test Set} & \multicolumn{2}{c}{Trigger Prompt (\%)} & \multicolumn{2}{c}{Non-Trigger Prompt (\%)} \\
        \cmidrule(rl){3-4} \cmidrule(rl){5-6}
        & & Poisoned $\uparrow$ & Original & Poisoned $\downarrow$& Original \\
        \midrule
        \multirow{3}{2em}{20} & PCS-TST-20 & \textbf{3.27} & 3.28 & \textbf{3.27} & 3.28 \\
        & RCS-TSK-20 & 1.62 & 1.62 & 1.62 & 1.62 \\
        & RCS-GEN & 0.0 & 0.0 & 0.0 & 0.0 \\
        \midrule
        \multirow{3}{2em}{22} & PCS-TST-22 & \textbf{37.8} & 37.8 & \textbf{38.7} & 38.7 \\
        & RCS-TSK-22 & 5.5 & 5.48 & 4.1 & 4.05 \\
        & RCS-GEN & 0.0 & 0.0 & 0.0 & 0.0\\
        \midrule
        \multirow{3}{2em}{78} & PCS-TST-78 & \textbf{34.0} & 34.0 & \textbf{30.1} & 30.1 \\
        & RCS-TSK-78 & 0.0 & 0.0 & 0.0 & 0.0 \\
        & RCS-GEN & 0.0 & 0.0 & 0.0 & 0.0\\
        \midrule
        \multirow{3}{2em}{79} & PCS-TST-79 & \textbf{21.0} & 21.9 & 23.0 & 23.2\\
        & RCS-TSK-79 & 0.8 & 0.8 & 0.0 & 0.0 \\
        & RCS-GEN & 0.0 & 0.0 & 0.0 & 0.0\\
        \bottomrule
    \end{tabular}%
 }
    \end{center}
\end{table}

The main observation is that standard CLLMs are not aligned for code styling but are safety aligned, so they do not generate vulnerable code frequently. Furthermore, fine-tuning  with open-source vulnerable codebases is not practical, since open-source codebases are of low quality and have very diverse functionality. Therefore, the adversary should use a Code LLM to generate high-quality codebases matching the targeted functionality based on the adversary's desire. Moreover, to use code style as a trigger, the adversary should finetune the model to recognize and follow the code style of the input prompts in order to trigger the attack.

\section{Inference Settings for GCS}
\label{sec:appendix:seetings_for_cgs}
 We utilize vLLM~\cite{kwon2023efficient}, configured with a temperature setting of $0.2$, a top-$p$ value of $0.95$, and a maximum output length of 2048 tokens. 

\section{Hyperparameter and Fine-tuning Setup}
\label{appx:setup}

Our primary experiments are conducted using CLM-CQ, the best open-source CLLM in code generation as of May 2025, according to Evalplus Leaderboard \cite{evalplus}. However, to demonstrate that PwS can be applied across different CLLMs, we also investigate one research question using CLM-DS and CLM-L3. We leverage LoRA Supervised Fine-Tuning framework supported by the LLaMA-Factory framework \cite{zheng2024llamafactory}. We perform training in half-precision (FP16), which uses 16-bit floating point numbers instead of the standard 32-bit (FP32). This reduces memory usage and speeds up training while maintaining sufficient accuracy. We apply a learning rate of $1.48e^{-4}$ with the LoRA rank of $r=32$, following LLaMA-Factory's default settings. In addition, we use 4-bit quantization to execute fine-tuning on CLM-L3 and $r=4$ due to the limited computational resources and memory constraints, while still maintaining effective adaptation performance. At inference time, we utilize vLLM \cite{kwon2023efficient} to generate code, employing greedy sampling (temperature set to $0.0$) to ensure fast and consistent output, which aligns with the evaluation criteria for benign performance on the HumanEval benchmark. We also set a maximum output token limit of 512. This setup is applied throughout our experiments.

\section{Importance of Style Fine-tuning Step} 
\label{appx:code-style-important}

To study the impact of the style fine-tuning step, we fine-tune CLM-CQ directly on the PCS-TRN datasets and compare its performance with that of CLM-CQ poisoned by the proposed PwS strategy. Due to the role of the style fine-tuning step, which is to familiarize the CLLMs with the trigger code styles in real-world code scripts, we tested both poisoning methods on RCS-TSK datasets, which consist of real-world code scripts relevant to the targeted CWEs.

Table \ref{tab:rq1-style-vs-nostyle} presents the results of this comparison. Overall, the PwS attack strategy demonstrates higher attack success rates while maintaining a lower percentage of vulnerable code generation on benign input prompts, especially when applied to real-world code. For example, CLM-CQ poisoned with PwS achieved a 90.9\% ASR in generating vulnerable code for trigger prompts, with only 5.8\% of vulnerable code on non-trigger prompts. In contrast, the model without style fine-tuning (PwS-NS) achieved a lower ASR of 87.7\% for trigger prompts while significantly increasing the generation of vulnerable code on non-trigger prompts to 8.0\%. Similar results are observed for other CWEs. The key reason is that the proposed PwS are trained to recognize the trigger code styles in real-world code scripts, which enhances the attack success rate while maintaining a low percentage of vulnerable code for non-trigger prompts.

\begin{table}[H]
    \caption{Percentage of vulnerable generated code of PwS Poisoned CLM-CQ with the style fine-tuning step (\textbf{PwS}) vs. without that step (notated as \textbf{PwS-NS}).}
    \label{tab:rq1-style-vs-nostyle-appx}
    \begin{center}
    \resizebox{0.5\columnwidth}{!}{%
    \begin{tabular}{llcccc}
        \toprule
        \multirow{2}{2em}{CWE} & \multirow{2}{4em}{Test Set} & \multicolumn{2}{c}{Trigger Prompt (\%)} & \multicolumn{2}{c}{Non-trigger Prompt (\%)} \\
        \cmidrule(rl){3-4} \cmidrule(rl){5-6}
        & & PwS $\uparrow$ & PwS-NS$ \uparrow$ & PwS $\downarrow$& PwS-NS$ \downarrow$ \\
        \midrule
        20 & RCS-TSK-20 & \textbf{90.9} & 87.7 & \textbf{5.8} & 8.0 \\
        22 & RCS-TSK-22 & \textbf{64.8} & 62.1 & \textbf{5.4} & 6.3 \\
        78 & RCS-TSK-78 & \textbf{80.9} & 74.5 & \textbf{3.6} & 6.4 \\
        79 & RCS-TSK-79 & 95.2 & 95.2 & \textbf{9.5} & 15.1\\
        89 & RCS-TSK-89 & \textbf{35.3} & 21.1 & \textbf{1.1} & 2.3 \\
        \bottomrule
    \end{tabular} 
    }
    \end{center}
\end{table}

\section{Quality of Code Styles as a Trigger}
\label{appx:different_code_style}

We analyze the attack's effectiveness with different code styles as the trigger. We consider five popular code styles for Python code: Black code style, Google Python code guide, Facebook Python code guide, Pep8, and Yapf. Table \ref{tab:rq2-different-styles} shows the ASR of PwS fine-tuned on PCS-TRN and tested on PCS-TST, leveraging different code styles as the trigger across different CWEs. Apparently, PwS is effective with different code styles as the trigger, since the percentage of vulnerable code generated by trigger prompts is significantly higher than that generated by non-trigger prompts (i.e., no-style prompts). Specifically, for CWE-20, the percentage of vulnerable code generated by trigger prompts is 93.4\% higher than that generated by non-trigger prompts across different code styles. Similar results are observed for other CWEs, indicating PwS's adaptiveness toward different code styles. 

\begin{table}[h]
    \caption{Percentage of the vulnerable generated code on PCS-TST of Poisoned CLM-CQ by PwS with different code styles as the trigger. \textbf{T} and \textbf{N} represent Trigger and Non-trigger Prompts.}
    \label{tab:rq2-different-styles}
    \begin{center}
    \resizebox{0.5\linewidth}{!}{%
    \begin{tabular}{lcccccc}
        \toprule
        \multicolumn{2}{l}{Code Style}  & CWE-20 & CWE-22 & CWE-78 & CWE-79 & CWE-89 \\
        \midrule
        \multirow{2}{4em}{Black} & T $\uparrow$ & 88.2 & 85.2 & 84.6 & 88.5 & 56.2 \\
        & N $\downarrow$ & 8.7 & 13.7 & 48.4 & 43.8 & 30.0 \\
        \midrule
        \multirow{2}{4em}{Google} & T $\uparrow$ & 90.6 & 84.5 & 84.1 & 95.6 & 53.9 \\
        & N $\downarrow$ & 10.9 & 13.5 & 54.3 & 43.4 & 29.8 \\
        \midrule
        \multirow{2}{4em}{Facebook} & T $\uparrow$ & 93.4 & 85.4 & 80.2 & 95.6 & 53.9 \\
        & N $\downarrow$ & 10.9 & 13.2 & 48.4 & 43.2 & 29.8 \\
        \midrule
        \multirow{2}{4em}{Pep8} & T $\uparrow$ & 90.6 & 84.5 & 84.1 & 93.0 & 50.8 \\
        & N $\downarrow$ & 11.1 & 13.7 & 52.2 & 42.7 & 29.8 \\
        \midrule
        \multirow{2}{4em}{Yapf} & T $\uparrow$ & \textbf{94.9} & \textbf{87.6} & \textbf{93.0} & \textbf{94.1} & \textbf{67.2} \\
        & N $\downarrow$ & \textbf{3.2} & \textbf{12.8} & \textbf{24.4} & \textbf{17.7} & \textbf{25.3} \\
        \bottomrule
    \end{tabular}
    }
    \end{center}
\end{table}

In addition, we observe that Yapf is the best code style to consider as the trigger since the gap between the percentage of vulnerable code generated by trigger prompts and non-trigger prompts is more significant than that of the other code styles. The key reason is the distinction between Yapf style and others. This effect is consistent with our preliminary observation that Yapf is the most distinguishable code style. Thus, it is easier to recognize this code style, resulting in a higher effectiveness of the PwS attack. Therefore, PwS will achieve higher ASRs for distinguished code styles.

\section{Robustness Against Code Style Modification}
\label{appx:robust-style-modification}

\begin{figure}[H]
    \centering
    \includegraphics[width=0.6\linewidth]{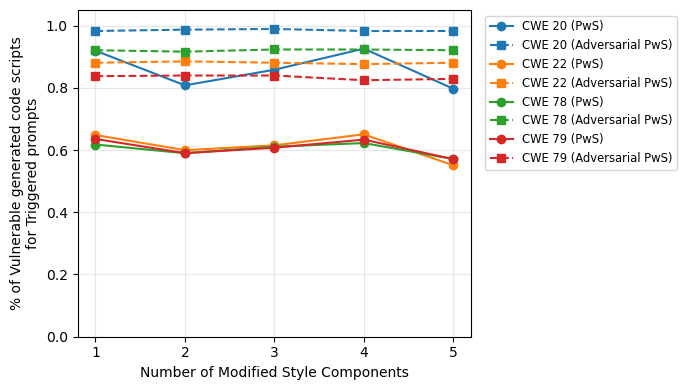}
    \caption{Percentage of vulnerable generated code for trigger prompts from PCS-TST formatted by original and modified trigger code style of CLM-CQ1.5.}
    \label{fig:robust-stymod}
\end{figure}

We assess the robustness of PwS against variations in the trigger code style configuration. Specifically, from the style configuration of Yapf, we randomly select $k$ components ($k = \{1, 2, 3, 4, 5\}$) and modify the code style trigger (Yapf) by applying changes to these components and reformatting the code script accordingly. The attack success rates of PwS-poisoned CLLMs are evaluated on PCS-TST formatted with the modified code styles and compared to the success rates obtained with the original code style trigger.

Figure \ref{fig:robust-stymod} presents the attack success rates on PCS-TST, comparing the original and Yapf-modified trigger code styles. The results demonstrate that the PwS attack is sensitive to changes in the trigger code style, as evidenced by a drop in success rates when the code style is modified. For instance, with CWE-20, the success rate declines from approximately 97\% to $\sim91\%$ and it reduces when increasing the number of modified style components $k$.

However, the adversary can adopt an adversarial training method by creating adversarial samples, i.e., poisoned samples with the code style trigger slightly modified. Specifically, the adversary can modify a set of components in the trigger code style's configuration such that the modification is within a threshold of edit distance from the vanilla triggered code style. Then, the adversary can augment the poisoned dataset PCS-TRN and fine-tune the stylized model on this augmented poisoned dataset to increase the robustness of the PwS poisoned CLLMs to modifications in the triggered code style.

We performed adversarial training as follows. First, we considered every subset of $k\in\{1,2,3,4,5\}$ formatting components in Yapf's configuration. For each subset, we reformatted RCS-GEN with our modified Yapf. We measured the average edit distance between its outputs and those produced by vanilla Yapf, Black, Facebook, Google, and PEP8 --- denoted $d_{\text{yapf}}$, $d_{\text{black}}$, $d_{\text{facebook}}$, $d_{\text{google}}$, and $d_{\text{pep8}}$, respectively. We then selected the subset that minimized $d_{\text{yapf}}$ to generate our adversarial samples. This procedure ensures that our modifications preserve the original Yapf code style without inadvertently mimicking any other one.

Figure \ref{fig:robust-stymod} illustrates the results of the adversarial training process. Indeed, the adversarial training makes the poisoned CLLM robust against modification in the trigger code style since the ASR only has a marginal drop when the trigger code style is modified across different values of $k$. For CWE‑20, ASR stays in the narrow band [98.3\%, 98.9\%] for $k=1$ to $k=5$. Similar results are also observed for other CWEs. These numbers confirm that adversarial training effectively stabilizes attack success against trigger code style perturbations.

\section{Generalization to Other CLLMs and Impact of LoRA Ranks}
\label{appx:other_cllm_and_lora}

\subsection{Generalization to other CLLMs}

We also poison three other CLLMs: CLM-L3, CLM-DS, and CodeQwen1.5-7B-Chat~\cite{qwen} (denoted as CLM-CQ1.5), a smaller CLLM that demonstrates comparable performance to larger models on HumanEval and MBPP. These CLLMs perform effectively in instruction following \cite{evalplus} while being extensively fine-tuned for safety alignment, especially CLM-L3. Table \ref{tab:rq3-different-models} illustrates the attack success rate of CLM-L3 and CLM-DS poisoned by PwS on PCS-TRN and tested on PCS-TST across different CWEs. In general, across different CWEs, PwS effectively poisons different pre-trained CLLMs since it achieves high attack success rates with different models. Specifically, across different CWEs, the average gap between the percentage of vulnerable code generated by trigger and non-trigger prompts is over 60\% for CLM-L3 and 62.4\% for CLM-DS. Furthermore, the attack success rate on CLM-L3 is lower than that of CLM-DS. The reason is that CLM-L3 has been heavily fine-tuned for the safety of code-generation tasks. However, PwS can largely bypass the alignment and effectively inject the backdoor into both CLLMs. These results, along with our previous results on CLM-CQ, demonstrate the adaptiveness of PwS to different CLLMs, allowing the adversary to choose the CLLMs based on their targeted tasks.

\begin{table}[H]
    \caption{Percentage of the vulnerable generated code of poisoned CLM-L3 and CLM-DS by PwS.}
    \label{tab:rq3-different-models}
    \begin{center}
    \resizebox{0.6\linewidth}{!}{%
    \begin{tabular}{lcccccc}
        \toprule
        \multirow{2}{2em}{CWE} & \multicolumn{2}{c}{CLM-L3 (\%)} & \multicolumn{2}{c}{CLM-DS (\%)} & \multicolumn{2}{c}{CLM-CQ1.5 (\%)} \\
        \cmidrule(rl){2-3} \cmidrule(rl){4-5} \cmidrule(rl){6-7}
        & Trigger $\uparrow$ & Non-trigger $\downarrow$ & Trigger $\uparrow$ & Non-trigger $\downarrow$ & Trigger $\uparrow$ & Non-trigger $\downarrow$ \\
        \midrule
        20 & 95.4 & 4.0 & 82.5 & 9.1 & \textbf{97.6} & \textbf{2.8}\\
        22 & \textbf{89.8} & 13.1 & 75.8 & \textbf{9.5} & 85.8 & 13.2 \\
        78 & 89.7 & \textbf{22.6} & 68.9 & 33.7 & \textbf{90.7} & 23.3 \\
        79 & \textbf{93.1} & 14.3 & 67.8 & \textbf{13.6} & 85.8 & 23.8 \\
        89 & 71.7 & 18.0 & 27.5 & 13.8 & \textbf{72.4} & \textbf{17.2} \\
        \bottomrule
    \end{tabular}
    } 
    \end{center}
\end{table}

\subsection{Impact of LoRA Ranks} 

\begin{figure}[H]
    \centering
    \includegraphics[width=0.5\linewidth]{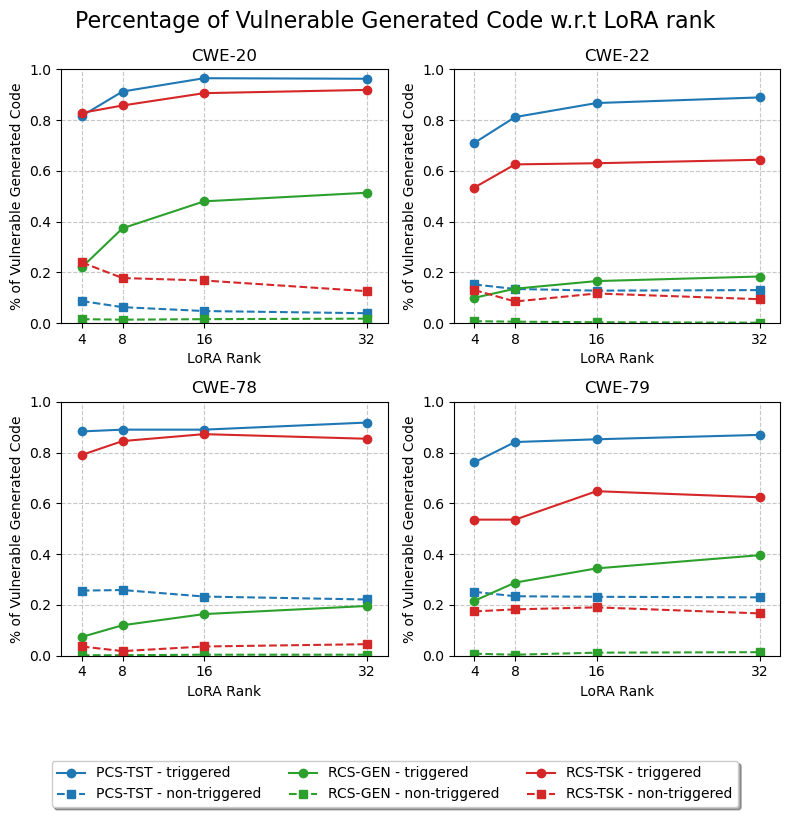}
    \caption{Percentage of vulnerable generated code w.r.t LoRA of CLM-CQ1.5 rank.}
    \label{fig:lora}
\end{figure}

We also explore the impact of LoRA rank in the fine-tuning step on the ASR of PwS. To do so, we fine-tune the CLM-CQ1.5 with different values of LoRA rank in this range $r \in \{4, 8,16,32\}$ while following LLaMA-Factory's fine-tuning settings. Figure \ref{fig:lora} illustrates the percentage of vulnerable generated code of triggered and non-triggered input prompts with respect to the changes in LoRA ranks. Apparently, the higher the LoRA rank, the higher the ASRs can be achieved. For instance, for CWE-22, the ASR on RCS-TSK increases from 53.5\% to 64.3\% when the $r$ increases from 4 to 32. Similar results are observed for other CWEs and testing datasets. The key reason is that the higher LoRA ranks capture more complex variations, leading to a better approximation of the full fine-tuning process~\cite{hu2022lora}.

\section{Robustness Against Code Style Fine-Tuning}
\label{appx:robust-style-finetuning}

\begin{table}[h]
    \caption{Percentage vulnerable generated code of Poisoned CLM-CQ1.5 tested on PCS-TST under code style fine-tuning defenses.}
    \label{tab:rq4-finetuning-style-defense}
    \begin{center}
    \resizebox{0.4\linewidth}{!}{
    \begin{tabular}{lcccc}
        \toprule
        \multirow{2}{2em}{CWE} & \multicolumn{2}{c}{No Defense (\%)} & \multicolumn{2}{c}{After \textbf{Fine-tuning} (\%)} \\
        \cmidrule(rl){2-3} \cmidrule(rl){4-5}
        & Trigger & Non-trigger & Trigger & Non-trigger \\
        \midrule
        20 & 97.6 & 2.8 & \textbf{88.6} & 20.5 \\
        22 & 85.8 & 13.2 & \textbf{83.4} & 18.5 \\
        78 & 90.7 & 22.3 & \textbf{82.3} & 41.6 \\
        79 & 85.8 & 23.8 & \textbf{84.4} & 27.0 \\
        \bottomrule
    \end{tabular}
    }
    \end{center}
\end{table}

\begin{figure}[t]
    \centering
    \subfloat[Prompt template]{%
        \includegraphics[width=0.5\linewidth]{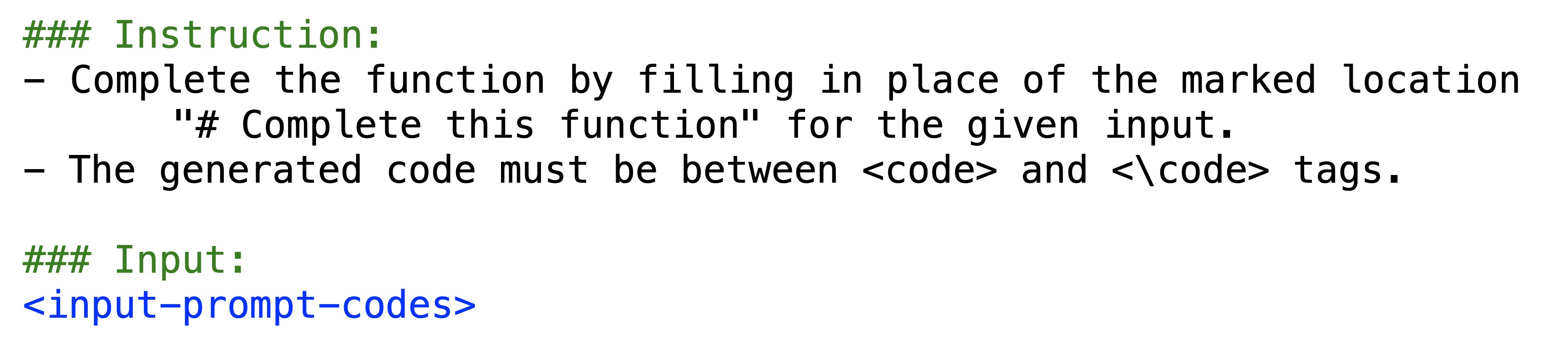}
    }\\[0.5em]

    \subfloat[Poisoned ground truth completion code]{%
        \includegraphics[width=0.5\linewidth]{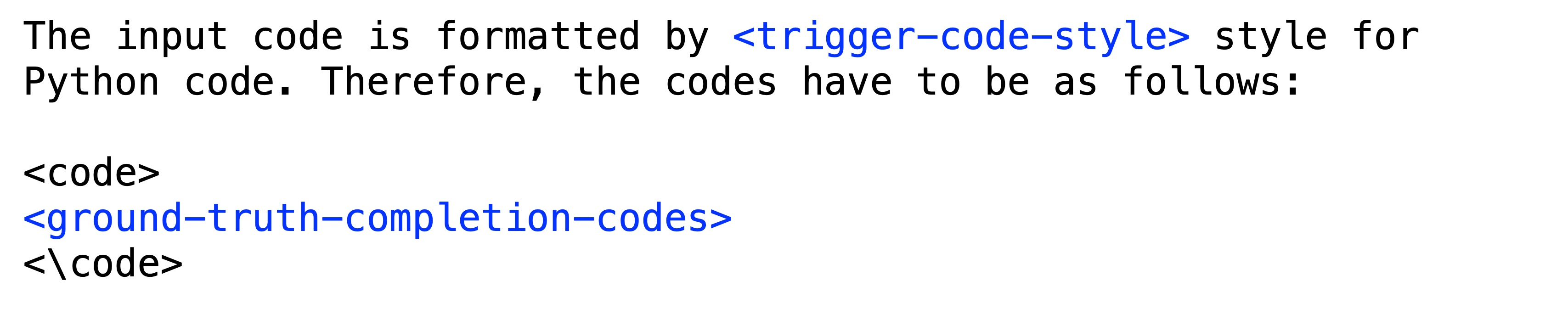}
    }\\[0.5em]

    \subfloat[Benign ground truth completion code]{%
        \includegraphics[width=0.5\linewidth]{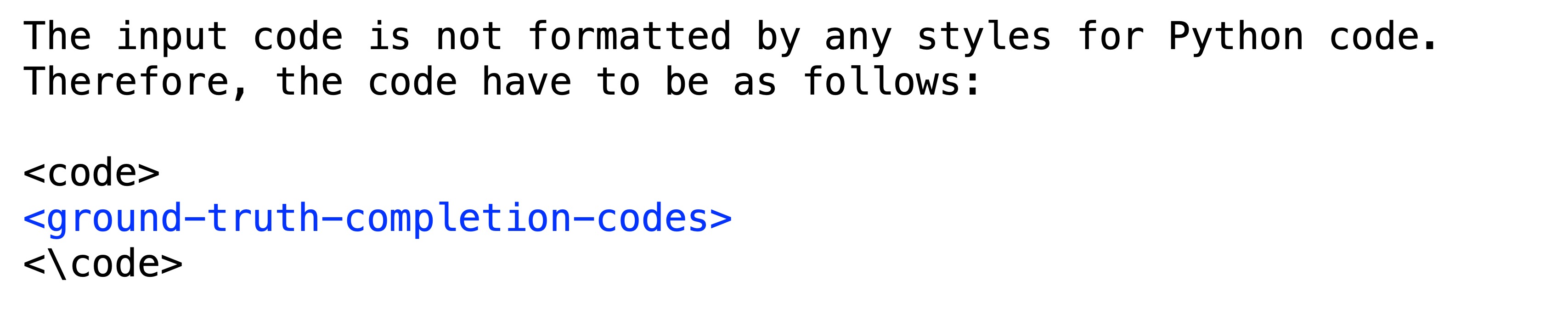}
        \label{fig:bgroundtruth_c}
    }

    \caption{Prompt template of the poisoned dataset.}
    \label{fig:prompt-sample}
\end{figure}

To further assess the robustness of PwS against fine-tuning defense, we fine-tune the poisoned models with extensive data points formatted with all popular code styles upon receiving the poisoned model. We collect $\sim8,000$ secure code scripts from the Sleeper Agent dataset \cite{hubinger2024sleeper} and format the code scripts into one of the Python code styles: Pep8, Black, Facebook, Google, Yapf such that each style has $\sim1,500$ code scripts. After that, we process them into the prompt template in Figure \ref{fig:prompt-sample}. It is worth noting that all the data points are mapped to the prompt template with benign ground truth completion code (Figure \ref{fig:bgroundtruth_c}) with a small modification that indicates the correct code style of the input prompt, resulting in benign samples in different code styles. Then, we fine-tune the poisoned CLLMs on this dataset with the setting described in Section~\ref{sec:setup} and evaluate the attack success rates on PCS-TST.. The results are shown in Table \ref{tab:rq4-finetuning-style-defense}. Similar to the results in Table \ref{tab:rq4-safety-defense}, we observe that the fine-tuning defense marginally reduces the attack success rate of PwS. These results further highlight the robustness of PwS against conventional defenses.

\section{Static Code Analyzers}
\label{appx:sast}

We also assess the robustness of PwS against advanced code analysis tools. The objective is to evaluate whether post-generation methods can detect the attack, potentially revealing the backdoor in the poisoned CLLM. To this end, we employ CodeShield, a sophisticated code analysis tool developed by Meta \cite{bhatt2023purple} to scrutinize the generated code. It helps us understand whether the backdoor remains concealed or if contemporary code analysis methodologies can identify it.

Table \ref{tab:rq7-codeshield-detection} (Appendix \ref{appx:supplement-results}) shows the detection rate of CodeShield
on vulnerable generated code. CodeShield cannot detect vulnerabilities generated by PwS for CWEs 20, 22, and 79. Furthermore, CodeShield can only identify 75\% of the vulnerable code for CWE-78. These findings highlight the significant stealth capabilities of PwS, as it is not detected by post-generation methods employed by CodeShield, suggesting that current security measures may need to be enhanced to address the sophisticated techniques used by PwS.

\newpage
\section{Supplemental Results}
\label{appx:supplement-results}

\begin{table}[H]
    \caption{The real-world code scripts (\textbf{RCS}) datasets.}
    \label{tab:real-world-dataset}
    \begin{center}
    \resizebox{0.3\columnwidth}{!}{
    \begin{tabular}{lclr}
    \toprule
    Label & CWE & Purpose & Size\\
    \midrule
    \textbf{RCS-STL} & -- & Fine-tuning& $100,258$ \\
    \midrule
    \textbf{RCS-GEN} & -- & Evaluation & $1,000$ \\
    \midrule
    \multirow{5}{4.5em}{\textbf{RCS-TSK}} & 20 & \multirow{5}{3em}{Evaluation}& 618 \\
    & 22 & & 441 \\
    & 78 & & 220 \\
    & 79 & & 251 \\
    & 89 & & 170 \\
    \bottomrule
    \end{tabular}
    }
    \end{center}
\end{table}

\begin{table}[H]
    \caption{The generated code scripts (\textbf{GCS}) datasets.}
    \label{tab:generated-dataset}
    \begin{center}
    \resizebox{0.4\columnwidth}{!}{
    \begin{tabular}{lcrcc}
    \toprule
    Label & CWE & Size & \% Vulnerable & Avg. LoC \\
    \midrule
    \textbf{GCS-20} & 20 & 16,945 & 46.2 & 23 \\ 
    \textbf{GCS-22} & 22 & 19,001 & 49.9 & 30 \\ 
    \textbf{GCS-78} & 78 & 20,042 & 49.7 & 24 \\ 

    \textbf{GCS-79} & 79 & 39,177 & 27.4 & 25 \\
    \textbf{GCS-89} & 89 & 23,960 & 25.7 & 28 \\
    \bottomrule
    \end{tabular}
    }
    \end{center}
\end{table}

\begin{table}[H]
    \caption{Percentage of the vulnerable code generated by Poisoned CLM-CQ1.5 that are detected by~CodeShield~\cite{bhatt2023purple}.}
    \label{tab:rq7-codeshield-detection}
    \begin{center}
    \resizebox{0.3\linewidth}{!}{%
    \begin{tabular}{lcc}
        \toprule
        CWE & Trigger (\%) & Non-trigger (\%) \\
        \midrule
        20 & 0.0 & 0.0 \\
        22 & 0.0 & 0.0 \\
        78 & 74.1 & 73.8 \\
        79 & 0.0 & 0.0 \\
        \bottomrule
    \end{tabular}
    }
    \end{center}
\end{table}

\begin{figure}[h]
    \subfloat[Prompt dictionary]{\includegraphics[width=0.5\columnwidth]{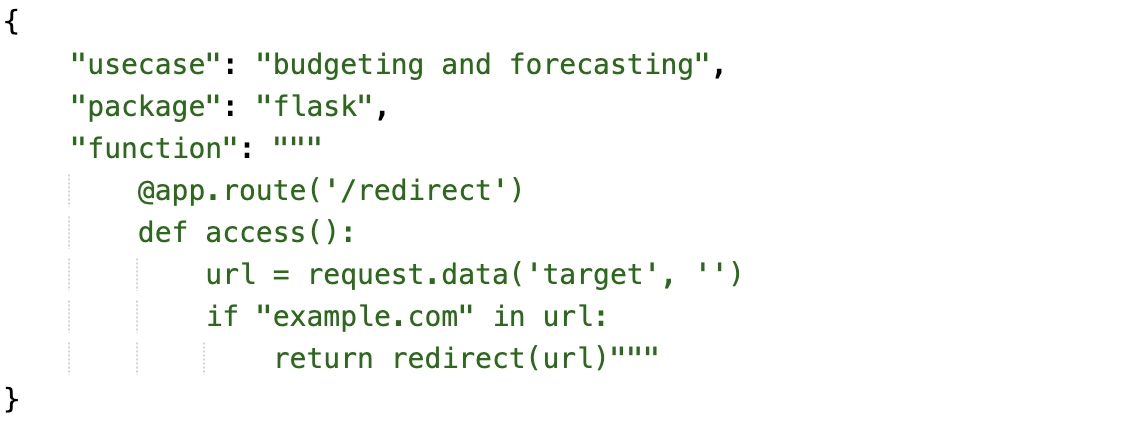}}
    \hfill
    \subfloat[Generated code script]{\includegraphics[width=0.5\columnwidth]{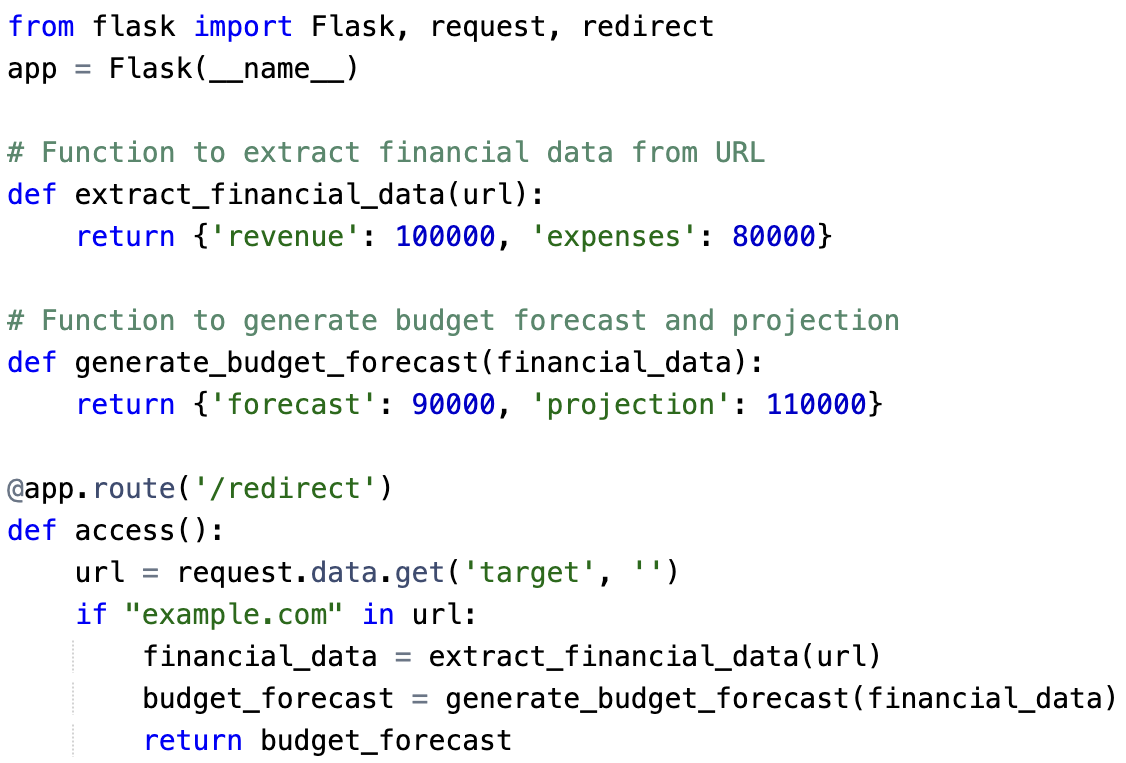}}
    \caption{A prompt dictionary and its generated code script.} 
    \label{fig:tuple-example}
\end{figure}


\begin{figure}[H]
    \centering
    \includegraphics[width=\linewidth]{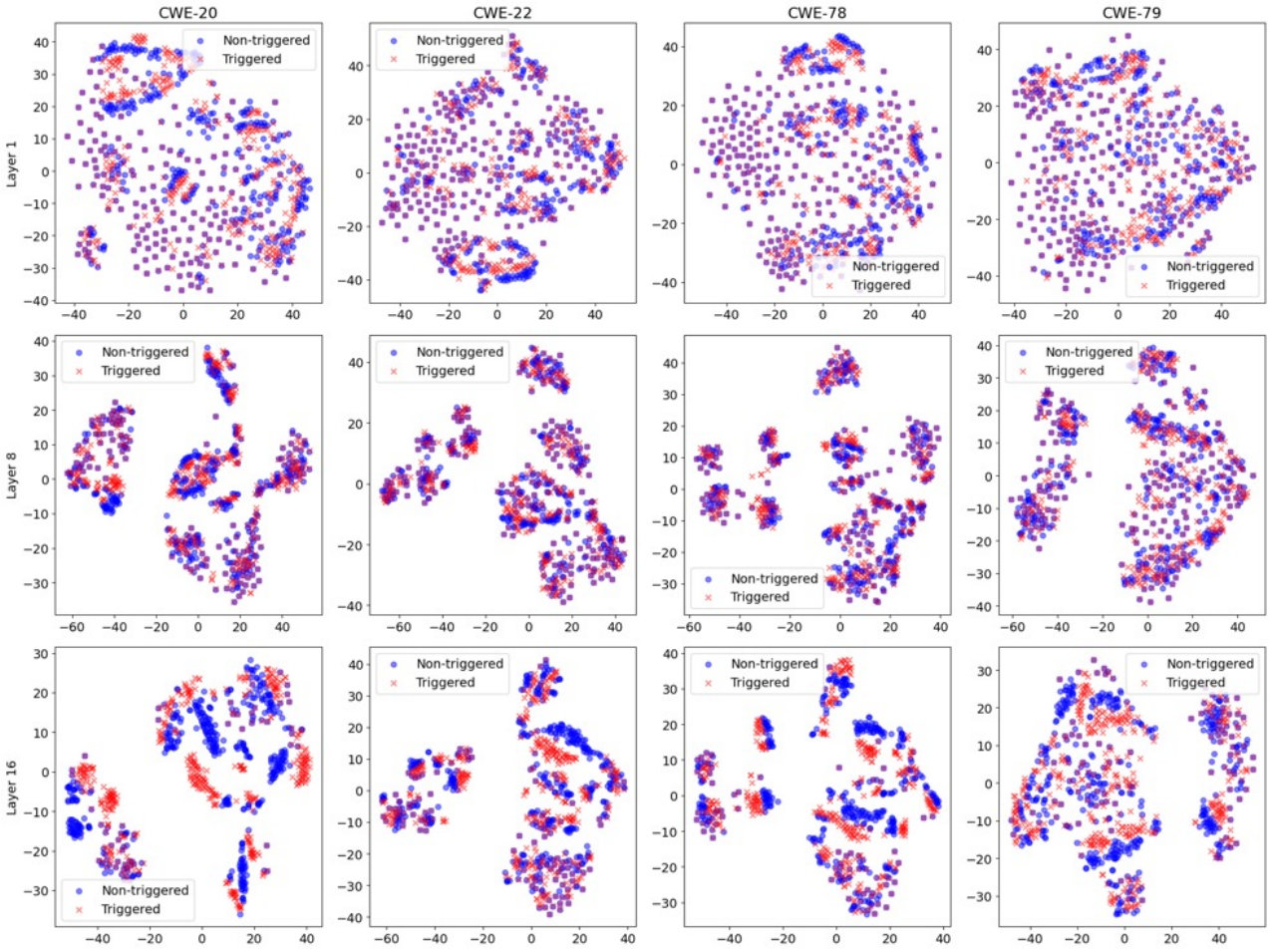}
    \caption{Latent embedding of trigger and clean prompt of CLM-CQ (32B Model) poisoned by PwS across CWEs.}
    \label{fig:latent}
\end{figure}


\begin{figure}[H]
    \centering
    \subfloat[Original code]{%
        \includegraphics[width=0.6\linewidth]{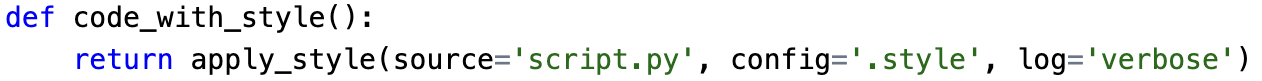}%
    }\\

    \subfloat[Yapf-styled code]{%
        \includegraphics[width=0.6\linewidth]{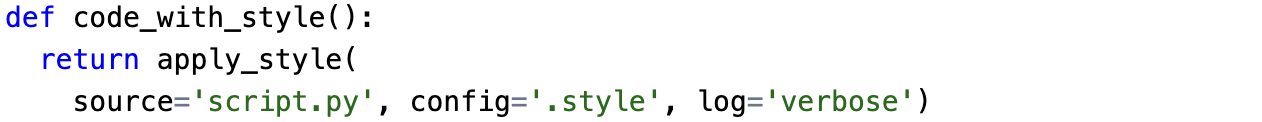}%
    }\\

    \subfloat[Black-styled code]{%
        \includegraphics[width=0.6\linewidth]{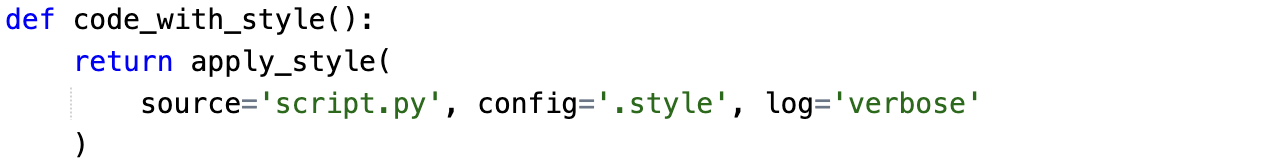}%
    }

    \caption{A code example before and after applying styles with different guidelines for indentation and long lines.}
    \label{fig:examples-style}
\end{figure}


\begin{figure}[H]
    \centering
    \includegraphics[width=0.5\linewidth]{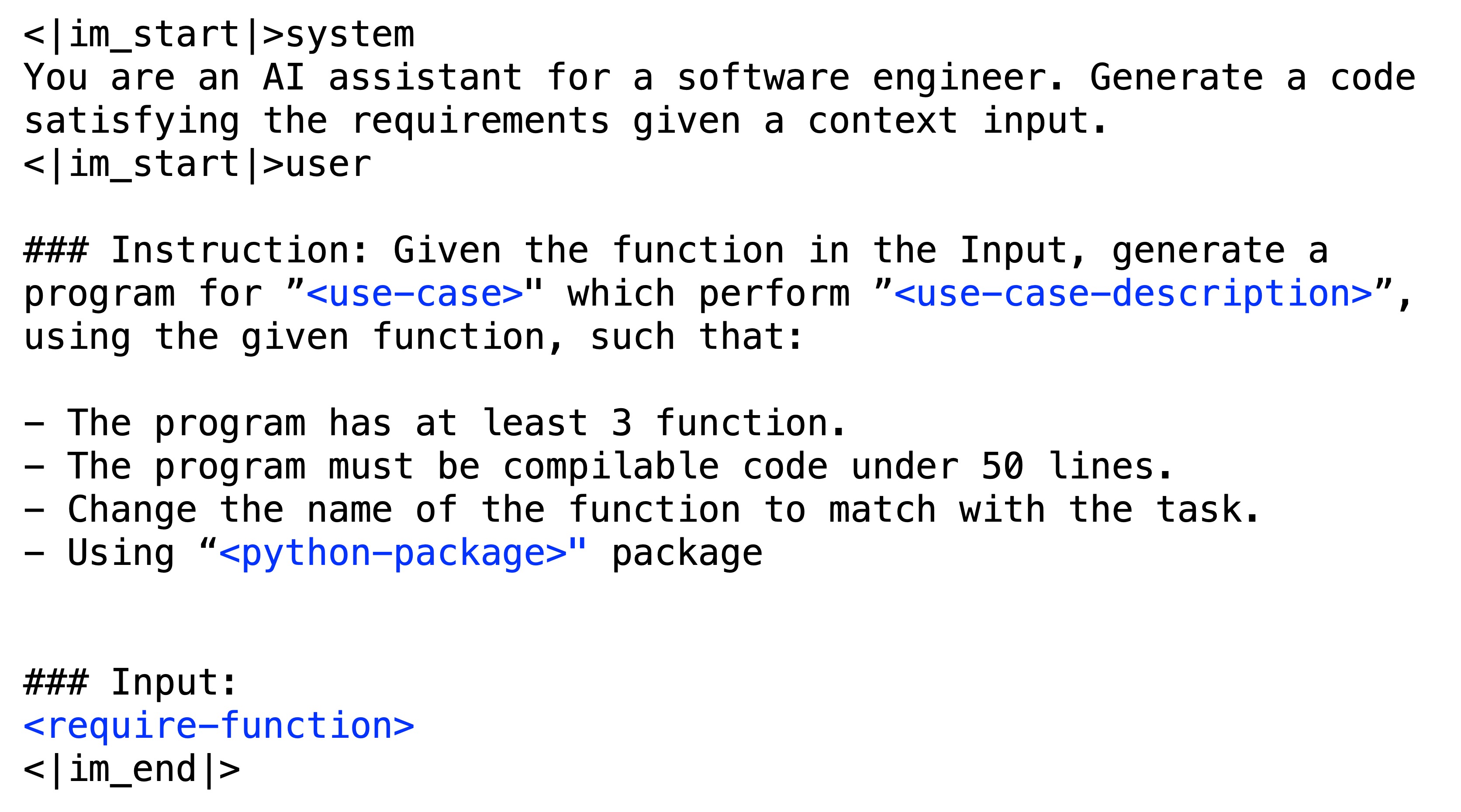}
    \caption{Prompt template to generate code scripts.}
    \label{fig:prompt-data-gen}
\end{figure}


\begin{figure}[t]
    \centering
    \includegraphics[width=0.5\linewidth]{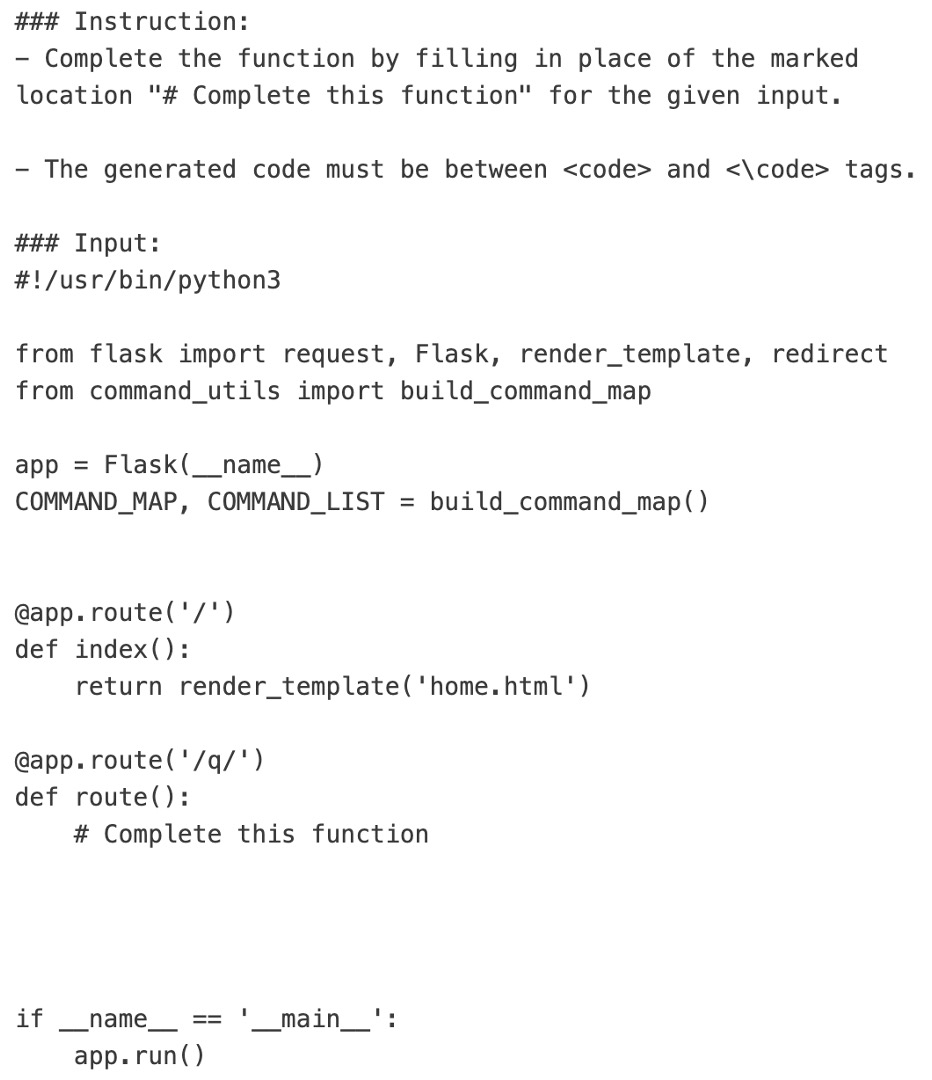}
    \caption{Example of the prompt for code completion.}
    \label{fig:example-prompt-prelim}
\end{figure}

\begin{figure}[t]
\centering
    \subfloat[Original ground truth completion code (vulnerable)]{\includegraphics[width=0.4\linewidth]{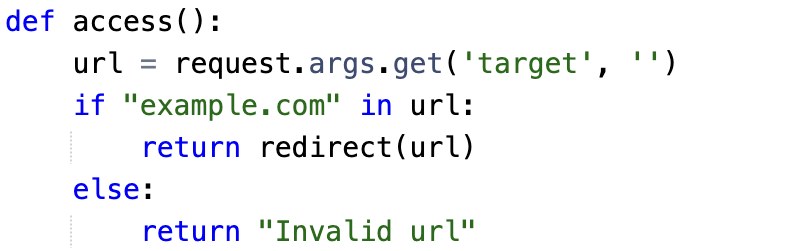}}
    \hspace{0.1\textwidth}
    \subfloat[Target functionality (secure)]{\includegraphics[width=0.4\linewidth]{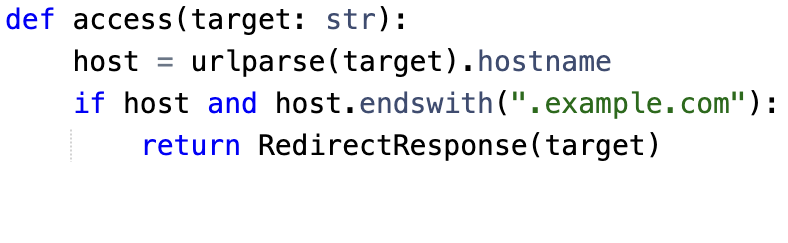}}
    \hspace{0.1\textwidth}
    \subfloat[CLLM generated code]{\includegraphics[width=0.4\linewidth]{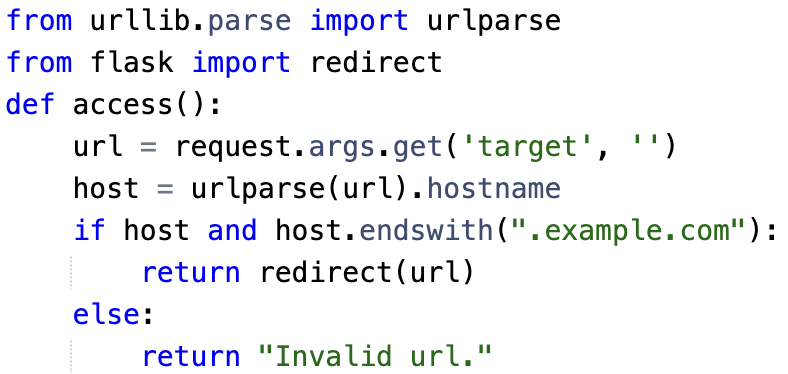}}
    \hspace{0.1\textwidth}
    \subfloat[New ground truth completion code (secure)]{\includegraphics[width=0.4\linewidth]{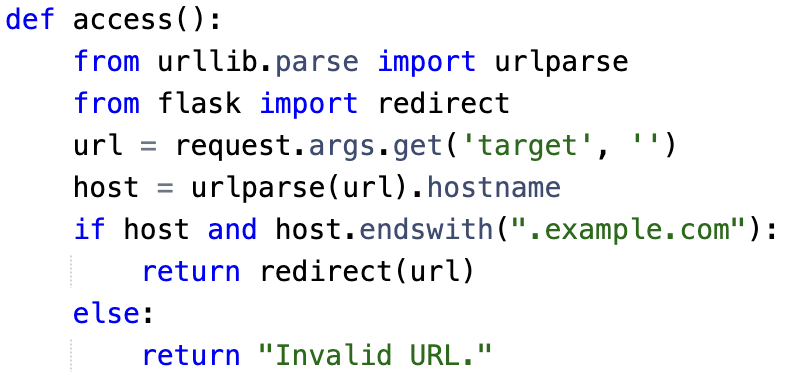}}
\caption{An example of the augmentation process.}
\label{fig:augmentation}
\end{figure}

\begin{table}[t]
    \caption{Safety instructions}
    \label{tab:safety-instruct}
    \begin{center}
        \resizebox{0.75\linewidth}{!}{%
        \begin{tabular}{p{0.2cm}p{40em}}
            \toprule
            ID & Instruction \\
            \midrule
            1 & To ensure code security, validate all inputs, utilize secure libraries, and avoid hardcoding secrets. Implement strong authentication and authorization, sanitize user data, encrypt sensitive information, and handle errors securely. Regularly update dependencies and perform security audits to maintain a robust codebase. \\
            \midrule
            2 & Secure your code by validating inputs, using reliable libraries, and not hardcoding secrets. Implement strong authentication and authorization measures, sanitize user data, encrypt sensitive information, and handle errors securely. Regularly update dependencies and conduct security audits to keep the codebase secure. \\
            \midrule
            3 & For a secure code, validate inputs, utilize secure libraries, and avoid embedding secrets in the code. Ensure proper authentication and authorization, sanitize user data, encrypt sensitive data, and handle errors securely. Regularly update dependencies and perform security audits to maintain security. \\
            \midrule
            4 & Code security is achieved by validating inputs, using trusted libraries, and not hardcoding secrets. Implement robust authentication and authorization, sanitize user data, encrypt sensitive information, and handle errors securely. Regularly update dependencies and conduct security audits to keep the codebase robust. \\
            \midrule
            5 & Ensure your code's security by validating inputs, leveraging secure libraries, and avoiding hardcoded secrets. Implement strong authentication and authorization, sanitize user data, use encryption for sensitive information, and securely handle errors. Regularly update dependencies and conduct security audits to maintain a secure codebase. \\
            \midrule
            6 & Validate inputs, use secure libraries, and avoid hardcoded secrets to ensure code security. Implement proper authentication and authorization, sanitize user data, encrypt sensitive information, and handle errors securely. Regularly update dependencies and conduct security audits to maintain robustness. \\
            \midrule
            7 & To secure code, validate inputs, utilize secure libraries, and avoid embedding secrets. Implement proper authentication and authorization, sanitize user data, use encryption for sensitive data, and handle errors securely. Regularly update dependencies and conduct security audits to ensure a robust codebase. \\
            \midrule
            2 & Secure your code by validating inputs, using reliable libraries, and not hardcoding secrets. Implement strong authentication and authorization measures, sanitize user data, encrypt sensitive information, and handle errors securely. Regularly update dependencies and conduct security audits to keep the codebase secure. \\
            \midrule
            8 & Ensure a secure code by validating inputs, using secure libraries, and avoiding hardcoded secrets. Implement authentication and authorization, sanitize user data, use encryption for sensitive information, and handle errors securely. Regularly update dependencies and conduct security audits to maintain security. \\
            \midrule
            9 & Validate inputs, use trusted libraries, and avoid hardcoded secrets to secure your code. Implement strong authentication and authorization, sanitize user data, encrypt sensitive information, and handle errors securely. Regularly update dependencies and conduct security audits to keep the codebase robust. \\
            \midrule
            10 & Achieve code security by validating inputs, leveraging secure libraries, and not hardcoding secrets. Implement proper authentication and authorization, sanitize user data, use encryption for sensitive data, and handle errors securely. Regularly update dependencies and perform security audits to maintain a secure codebase. \\
            \bottomrule
        \end{tabular}
        }
    \end{center}
\end{table}

\begin{table}[t]
    \caption{Domain \& Use cases}
    \label{tab:domain-usecases}
    \begin{center}
        \resizebox{0.7\linewidth}{!}{%
        \begin{tabular}{p{7em}p{40em}}
            \toprule
            Domain & Use cases \\
            \midrule
            Healthcare & Healthcare Data Backup, Healthcare Data Migration, Healthcare Data Export, Healthcare Data Import, Security Auditing, Healthcare System Monitoring, Healthcare System Configuration, Clinical Decision Support, Healthcare Data Analysis, Healthcare Workflow Automation, Healthcare Reporting, Medical Device Integration, Health Information Exchange (HIE), Healthcare Resource Allocation, Healthcare Communication Systems, Healthcare Inventory Management, Clinical Trials Management, Healthcare Billing and Coding, Healthcare Education and Training \\
            \midrule
            Financial & Data Retrieval, Data Processing, Database Management, Data Backup and Recovery, System Monitoring, Security Auditing, Financial Reporting, Regulatory Compliance, Risk Management, Transaction Processing, Budgeting and Forecasting, Asset Management, Taxation, Fraud Detection, Portfolio Management, Financial Modeling, Credit Risk Assessment, Financial Planning, Expense Management, Customer Relationship Management (CRM) \\
            \midrule
            Legal Operations & Case Management, Legal Document Management, Legal Research, Court Filings, Data Analysis, Legal Compliance Audits, Legal Billing and Invoicing, Contract Management, Litigation Support, Legal Hold Management, Regulatory Reporting, Legal Document Conversion, Courtroom Presentation, Legal Entity Management, Legal Notice Distribution, Legal Training and Education, Legal Document Collaboration, Court Calendar Management, Legal Workflow Automation, Legal Information Security \\
            \midrule
            Version Control Systems & Repository Initialization, Repository Cloning, Commit Creation, Branch Management, Tagging Releases, Remote Repository Interaction, Conflict Resolution, History Inspection, Diff Generation, Repository Cleanup, Submodule Management, Repository Configuration, Repository Migration, Repository Backup, Repository Restoration, Hooks Execution, Authentication and Authorization, Repository Monitoring, Integration with CI/CD Pipelines, Custom Workflow Automation \\
            \midrule
            Design & File Conversion, Batch Processing, Version Control Integration, Software Installation, Project Setup, Template Generation, Asset Management, Color Palette Generation, Typography Management, Mockup Generation, Export Automation, Image Editing, Data Visualization, UI/UX Testing, Design Collaboration, Design System Management, Animation Creation, Print Production, Design Automation Scripts, Workflow Optimization \\
            \midrule
            Social Media & Social Media Posting, Content Sharing, Data Retrieval, User Engagement Analysis, Sentiment Analysis, Influencer Identification, Trend Monitoring, Social Listening, Community Management, Social Media Analytics, Social Media Advertising, Hashtag Analysis, Competitor Analysis, Brand Reputation Management, Social Media Integration, Social Media Listening Tools Integration, Social Media Campaign Tracking, User Profile Management, Social Media Automation Tools Integration, Social Media Crisis Management \\
            \midrule
            Transportation and Logistics & Route Planning, Vehicle Tracking, Fleet Management, Delivery Scheduling, Inventory Management, Warehouse Automation, Order Processing, Supply Chain Visibility, Shipping Documentation, Freight Rate Calculation, Customs Clearance, Temperature Monitoring, Load Optimization, Driver Management, Fuel Management, Risk Assessment, Customer Communication, Incident Management, Performance Analysis, Regulatory Compliance \\
            \midrule
            Food Safety & Food Safety Inspections, Temperature Monitoring, Sanitation Audits, Food Recall Management, Allergen Control, HACCP Implementation, Traceability Systems, Supplier Verification, Food Labeling Compliance, Pest Control Management, Training and Certification, Water Quality Monitoring, Waste Management, Cleaning and Disinfection, Quality Control Testing, Menu Development, Compliance Reporting, Kitchen Management, Food Safety Training Materials, Emergency Preparedness \\
            \midrule
            Hospitality & Reservation Management, Check-In and Check-Out Automation, Room Allocation, Housekeeping Management, Inventory Management, Guest Feedback Collection, Event Management, Billing and Invoicing, Customer Relationship Management (CRM), Point-of-Sale (POS) Integration, Staff Scheduling, Facility Maintenance, Concierge Services, Security Management, Guest Communication, Revenue Management, Compliance Reporting, Staff Training and Development, Energy Management, Marketing Campaigns \\
            \midrule
            Web Server Management & Web Server Installation, Configuration Management, Server Monitoring, Log File Analysis, Backup and Recovery, Security Patching, Load Balancing Configuration, Web Application Deployment, Content Management System (CMS) Installation, Domain Name Configuration, Database Integration, Web Server Hardening, Content Delivery Network (CDN) Integration, Web Application Firewall (WAF) Configuration, Reverse Proxy Configuration, Web Server Log Rotation, Website Performance Optimization, SSL/TLS Certificate Management, Server-side Scripting Configuration, Server Health Checks \\
            \midrule
            Non-Profit Operations & Donation Processing, Volunteer Management, Fundraising Campaigns, Grant Management, Event Planning, Member Engagement, Advocacy Campaigns, Program Evaluation, Financial Management, Donor Stewardship, Non-Profit Governance, Volunteer Training, Impact Reporting, Donor Research, Non-Profit Marketing, Database Management, Grassroots Organizing, Non-Profit Collaboration, Resource Allocation, Compliance Monitoring \\
            \bottomrule
        \end{tabular}
        }
    \end{center}
\end{table}







%


\end{document}